\documentclass[3p,twocolumn,final]{elsarticle}

\usepackage{lineno,hyperref}
\usepackage{amsmath,esint}
\usepackage{amssymb,wasysym}
\usepackage{amsfonts}
\usepackage[noabbrev,capitalize]{cleveref}
\usepackage{overpic}
\usepackage{xcolor}
\usepackage[normalem]{ulem}
\definecolor{forestgreen}{rgb}{0.13, 0.72, 0.19}

\modulolinenumbers[5]

\makeatletter
\def\ps@pprintTitle{%
   \let\@oddhead\@empty
   \let\@evenhead\@empty
   \def\@oddfoot{\reset@font\hfil\thepage\hfil}
   \let\@evenfoot\@oddfoot
}
\makeatother

\newcommand{\lt}{\left}
\newcommand{\rt}{\right}

\begin{document}

\begin{frontmatter}

\title{Classification of chaotic time series with deep learning}

%% or include affiliations in footnotes:
\author[Oxford]{Nicolas Boull\'e\corref{mycorrespondingauthor}}
\ead{boulle@maths.ox.ac.uk}

\author[Oxford]{Vassilios Dallas\corref{mycorrespondingauthor}}
\ead{vassilios.dallas@gmail.com}

\author[Oxford]{Yuji Nakatsukasa}
\ead{nakatsukasa@maths.ox.ac.uk}

\author[CCFE]{D. Samaddar}
\ead{debasmita.samaddar@ukaea.uk}

\cortext[mycorrespondingauthor]{Corresponding author}

\address[Oxford]{Mathematical Institute, University of Oxford, Oxford, OX2 6GG, UK}
\address[CCFE]{United Kingdom Atomic Energy Authority, Culham Centre for Fusion Energy,\\ Culham Science Centre, Abingdon, Oxon, OX14 3DB, UK}

\begin{abstract}
We use standard deep neural networks to classify univariate time series generated by discrete and continuous dynamical systems based on their chaotic or non-chaotic behaviour. Our approach to circumvent the lack of precise models for some of the most challenging real-life applications is to train different neural networks on a data set from a dynamical system with a basic or low-dimensional phase space and then use these networks to classify univariate time series of a dynamical system with more intricate or high-dimensional phase space. We illustrate this generalisation approach using the logistic map, the sine-circle map, the Lorenz system, and the Kuramoto--Sivashinsky equation. We observe that a convolutional neural network without batch normalization layers outperforms state-of-the-art neural networks for time series classification and is able to generalise and classify time series as chaotic or not with high accuracy.
\end{abstract}

\begin{keyword}
Dynamical systems\sep Chaos\sep Deep learning\sep Time series\sep Classification
\end{keyword}

\end{frontmatter}

% Comment to remove line number
%\linenumbers

\section{Introduction}

% Machine Learning for Dynamical Systems: how to analyze dynamical systems on the basis of observed data rather than attempt to study them analytically.

% While models are very precise for many processes, for some of the most challenging applications of dynamical systems the development of such models is notably difficult.

% For infinite dimensional dynamical systems is not possible to be able to explore the whole phase space of the system

Data and in particular time series are generated from numerous observations and experiments across different scientific fields such as atmospheric and oceanic sciences for climate predictions, nuclear fusion for control and safety, biology and medicine for diagnosis. Fourier transforms, radial basis functions approximation and standard numerical techniques have been extensively applied to perform short and long term predictions of chaotic time series~\cite{casdagli1989nonlinear,lai2003recent,mukherjee1997nonlinear,sugihara1994nonlinear}. On the other hand, the spectacular success of machine learning and deep learning techniques to image classification~\cite{krizhevsky2012imagenet,lecun1998handbook}, which have recently surpassed human-level performance on the ImageNet data set~\cite{he2015delving}, has inspired the development of neural network techniques for time series forecasting~\cite{elsner1992nonlinear,kuremoto2014time} and classification~\cite{wang2017time}. Recently, deep learning approaches have been used to solve partial differential equations in high dimensions~\cite{han2018solving,sirignano2018dgm,weinan2018deep} and identify hidden physics models from experimental data~\cite{raissi2018deep,raissi2018hidden,raissi2019physics,rudy2017data}.  

The size of the data sets is often large and analysing these time series represents a huge computational challenge and interest nowadays. For some of the most challenging real-life applications a precise dynamical system is unknown, which makes the identification of the different dynamical regimes impossible. In that spirit, machine learning has been recently employed by Pathak et al.~\cite{pathak2018model,pathak2017using} to perform model-free predictions of chaotic dynamical systems. Moreover, deep learning requires a large data set to adequately train the artificial neural network, which might not be available in some cases due to the infinite dimensional phase space of the system or experimental constraints.

In this paper, we address the aforementioned challenges by considering the following classification problem: given a univariate time series generated by a discrete or continuous dynamical system, can we determine whether the time series has a chaotic or non-chaotic behaviour? We choose to train neural networks on a different set than the testing set of interest in order to assess the ability of the machine learning algorithms to generalise the classification of univariate time series of a dynamical system with a basic or low-dimensional phase space to a more intricate or high-dimensional one. Our aim is to demonstrate the generalisation ability of neural networks in this classification problem using standard deep learning models. The main challenge here is to learn the chaotic features of a training set, whose chaotic behaviour can be determined a priori using measures from dynamical systems theory, without overfitting, and generalise on a testing data set, which comes from another dynamical system, whose behaviour is different.

The paper is organised as follows. We briefly describe five different neural networks architectures for time series classification in \cref{sec_Net_TSC}. Then, in \cref{sec_discrete_res}, we classify signals generated by discrete dynamical systems and compare the accuracy of the neural networks. Finally, \cref{sec_continuous_res} consists of the classification of time series generated by the Lorenz system and the Kuramoto--Sivashinsky equation.

\section{Neural networks for time series classification} \label{sec_Net_TSC}

Time series classification is one of the most challenging problems in machine learning~\cite{esling2012time} with a wide range of applications in human activity recognition~\cite{nweke2018deep}, acoustic scene classification~\cite{nwe2017convolutional}, and cybersecurity~\cite{susto2018time}. In this section, we describe five different architectures that we have considered for classifying time series generated by discrete and continuous dynamical systems.

First, we consider a simple shallow neural network (see \cref{sec_SNN}) in order to compare it with state-of-the-art classifiers. The multilayer perceptron~\cite{wang2017time}, presented in \cref{sec_MLP}, is chosen because it is a fully connected deep neural network, which does not rely on convolutional layers. Convolutional neural networks have first been introduced to perform handwritten digit~\cite{lecun1989backpropagation} and have been successfully applied to images and time series~\cite{krizhevsky2012imagenet, lecun1998handbook,lecun1998gradient}. The next two neural networks (see \cref{sec_FCN} and \ref{sec_ResNet}) are then based on convolutional layers and were chosen on the basis that they achieve the best performance on standard time series datasets according to the recent review by Fawaz et al.~\cite{fawaz2019deep}. Finally in \cref{sec_convnet} we consider a convolutional neural network for time series classification with a large kernel size to decrease the computational expense.

In what follows, the input of the neural network is a univariate time series $X$ of size $T$ (in practice we take $T$ to be one thousand). Each of the time series is assigned a class label that we want to recover using the different neural networks: Class $NC$ corresponds to a non-chaotic time series while Class $C$ corresponds to a chaotic time series. The neural networks presented in this section end with a softmax layer, which outputs a vector of probabilities $[p_{NC},p_C]$, where $p_{NC}$ is the probability that the input time series is non-chaotic and $p_C=1-p_{NC}$ is the probability that it is chaotic. A time series is classified as non-chaotic if $p_{NC}\geq 0.5$ and chaotic otherwise. The performance of the neural networks is then assessed using the %categorical 
classification accuracy defined as
\[ \text{Accuracy} = \frac{T_{NC}+T_C}{T_{NC} + T_C + F_{NC} + F_C} \times 100,\]
where $T_{NC}$ = true non-chaotic predictions, $T_C$ = true chaotic predictions, $F_{NC}$ = false non-chaotic predictions, and $F_C$ = false chaotic predictions.

Analysing our results with other metrics than accuracy can be important particularly when the testing data sets are not balanced. For this reason we also considered metrics such as precision, recall~\cite[Chapt.~9]{olson2008advanced}, and balanced accuracy~\cite{brodersen2010balanced}. However, we saw that our findings were not influenced by these metrics even though their values change. Thus, we choose to report only the values of the classification accuracy which are not better or worse than the other metrics.

\subsection{Shallow neural network}\label{sec_SNN}

The first type of networks considered in this section is shallow neural networks (ShallowNet), which are simple and efficient networks for fitting functions and perform pattern recognition. These networks differ from deep neural network since they usually contain only one or two hidden layers. We use MATLAB's \texttt{patternnet} command from the Deep Learning Toolbox to define a network with one hidden layer, containing one hundred neurons, with the sigmoid as activation function. The network is trained with the Scaled Conjugate Gradient algorithm~\cite{moller1993scaled} using MATLAB's default settings.

\subsection{Multi layer perceptrons} \label{sec_MLP}
Multilayer perceptrons (MLP) are standard deep neural network architectures in the field of machine learning and essentially consist of fully connected layers separated by a nonlinear activation function. Wang, Weizhong and Oates~\cite{wang2017time} use a structure of three hidden layers of five hundred neurons followed by a rectified linear unit (ReLU) to perform time series classification (a Python implementation using TensorFlow is available in~\cite{Wang2019}). Moreover, a dropout is added at the input, hidden and softmax layers with rates $\{0.1,0.2,0.3\}$, respectively. The network is trained on 10 epochs with a batch size of 32 using Adam's algorithm~\cite{kingma2014adam}. We use the same parameters and optimisation algorithm to train the following three convolutional neural networks.

\subsection{Fully convolutional neural network} \label{sec_FCN}
The fully convolutional neural network (FCN) architecture considered in~\cite{wang2017time} is a succession of three convolutional blocks, followed by a global averaged pooling layer~\cite{lin2013network} and a softmax layer. The first (resp. second, third) convolutional block that we consider is composed of a convolution layer of kernel size $8$ (resp. $5$, $3$) with $64$ (resp. $128$, $64$) feature channels, a batch normalization layer~\cite{ioffe2015batch} and a ReLU activation layer. The fully convolutional neural network studied by~\cite{wang2017time} is implemented in~\cite{Wang2019}.

\subsection{Residual network} \label{sec_ResNet}

The last network considered by Wang, Weizhong and Oates~\cite{wang2017time} in the context of time series classification is a residual network (ResNet). Residual networks are examples of very deep neural network and are designed by stacking the convolutional blocks arising in the FCN (see \cref{sec_FCN}). Then, the ResNet is created by assembling three blocks of the FCN to generate a residual block. Three residual blocks, with $\{64,128,128\}$ respective number of feature channels, are then stacked and followed by a global average pooling layer and a softmax layer to output the classification of the different input time series. Reference \cite{Wang2019} provides a practical Python implementation.

\subsection{Large kernel convolutional neural network} \label{sec_convnet}

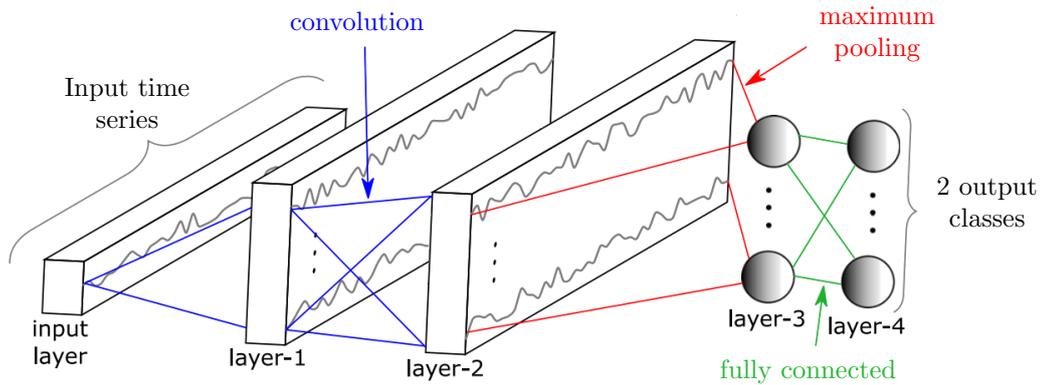
\begin{figure*}[htbp]
\centering
\begin{overpic}[width=0.8\textwidth]{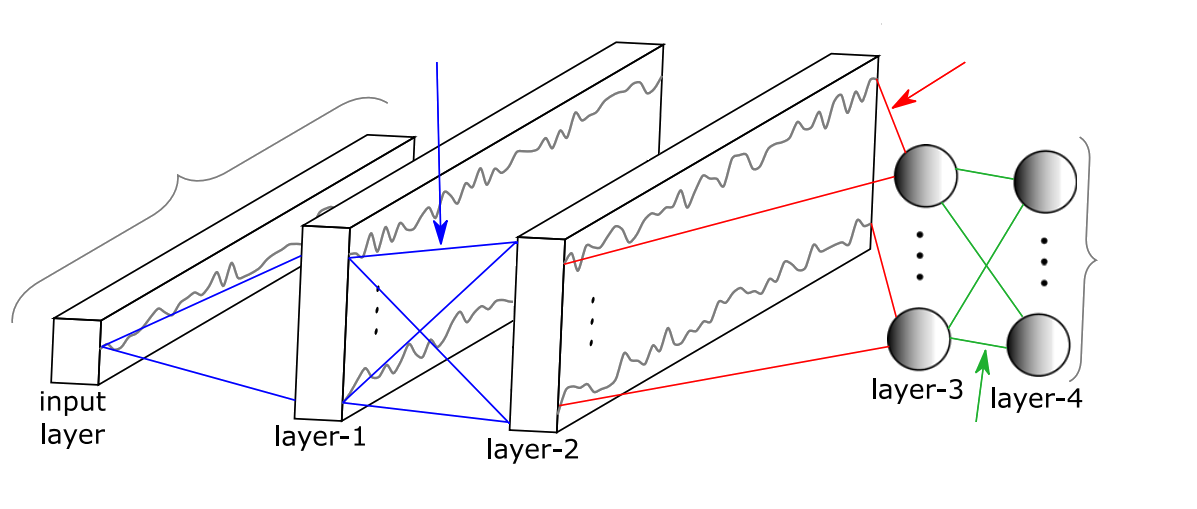}
\put(5,31){\parbox{.8in}{\centering Input time series}}
\put(29,39){\textcolor{blue}{convolution}}
\put(80,38){\parbox{.8in}{\centering \textcolor{red}{maximum pooling}}}
\put(91,21){\parbox{.8in}{\centering $2$ output classes}}
\put(72,4){\textcolor{forestgreen}{fully connected}}
\end{overpic}
\caption{The large kernel convolutional neural network (LKCNN) architecture for time series classification. Figure adapted from~\cite{fawaz2019deep}.}
\label{fig_conv1d}
\end{figure*}

We consider a standard convolution neural network with large kernel size (LKCNN). The network is composed of two convolutional layers with five feature channels and kernel size of a hundred, followed by ReLU activation, a maximum pooling layer with a pool size of two, a flatten layer, and two fully connected layers of respective size one hundred and two. The ReLU activation function is chosen because it is easy to optimise due to its piecewise linearity~\cite[Chap.~6]{goodfellow2016deep}. Moreover, a dropping out unit (dropout) with rate $0.5$ is added after the maximum pooling layer to improve the generalisation ability of the network~\cite{srivastava2014dropout}. The architecture of the network is shown in \cref{fig_conv1d}.

Standard implementations of convolutional neural networks usually consider a larger number of feature channels and a much smaller kernel size~\cite{wang2017time,fawaz2019deep}. However, we observed that the classification accuracy of this network was not affected by the kernel size and decreased as we increased the number of feature channels (see \ref{appendix_a}). Thus, we choose to use a large kernel size and a small number of feature channels in order to reduce the number of trainable parameters and thus reduce the computational expense of this network. Similarly, we verified that increasing the number of convolutional layers from two to three or four does not significantly improve the performance of LKCNN overall.

The main difference that determines the classification accuracy between LKCNN, FCN, and ResNet is that FCN and ResNet have batch normalization layers. Batch normalization was introduced to speed up the optimisation algorithm that occurs at the training phase by normalising the training input data at the internal layers~\cite{ioffe2015batch}.

In this work, we study the generalisation ability from a training set to a testing set with time series that span different range of values. For this reason, we normalize both the training set and the testing set (see details in the following sections) so that the values vary within the same range. Note that the scaling and shifting parameters of the batch normalisation layers are determined in the training phase \cite{ioffe2015batch}. Therefore, applying a batch normalisation to a data set with different mean and variance would be wrong. In our problem this happens with the testing data, which can have different mean and variance from the training data after the convolutional layers. This explains the lack of performance of the FCN and ResNet on the testing data sets as we will show in the following sections.

\section{Discrete dynamical systems} \label{sec_discrete_res}

\begin{figure}[htbp]
\centering
\begin{minipage}{0.49\textwidth} 
\begin{overpic}[width=\textwidth]{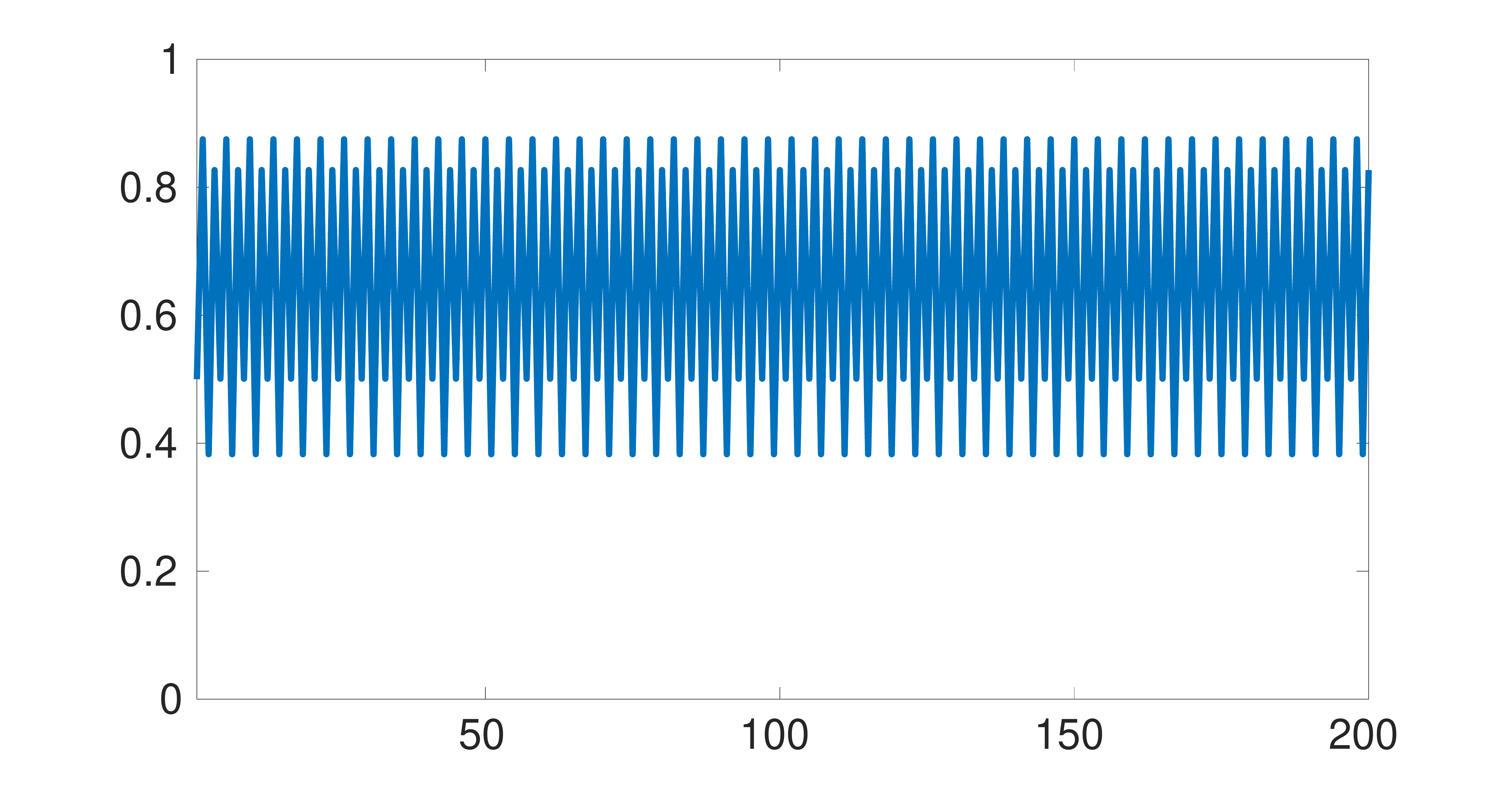}
\put(51,-1){$n$}
\put(1,26){$x_n$}
\end{overpic}
\end{minipage}\\
\vspace{0.2cm}
\begin{minipage}{0.49\textwidth} 
\begin{overpic}[width=\textwidth]{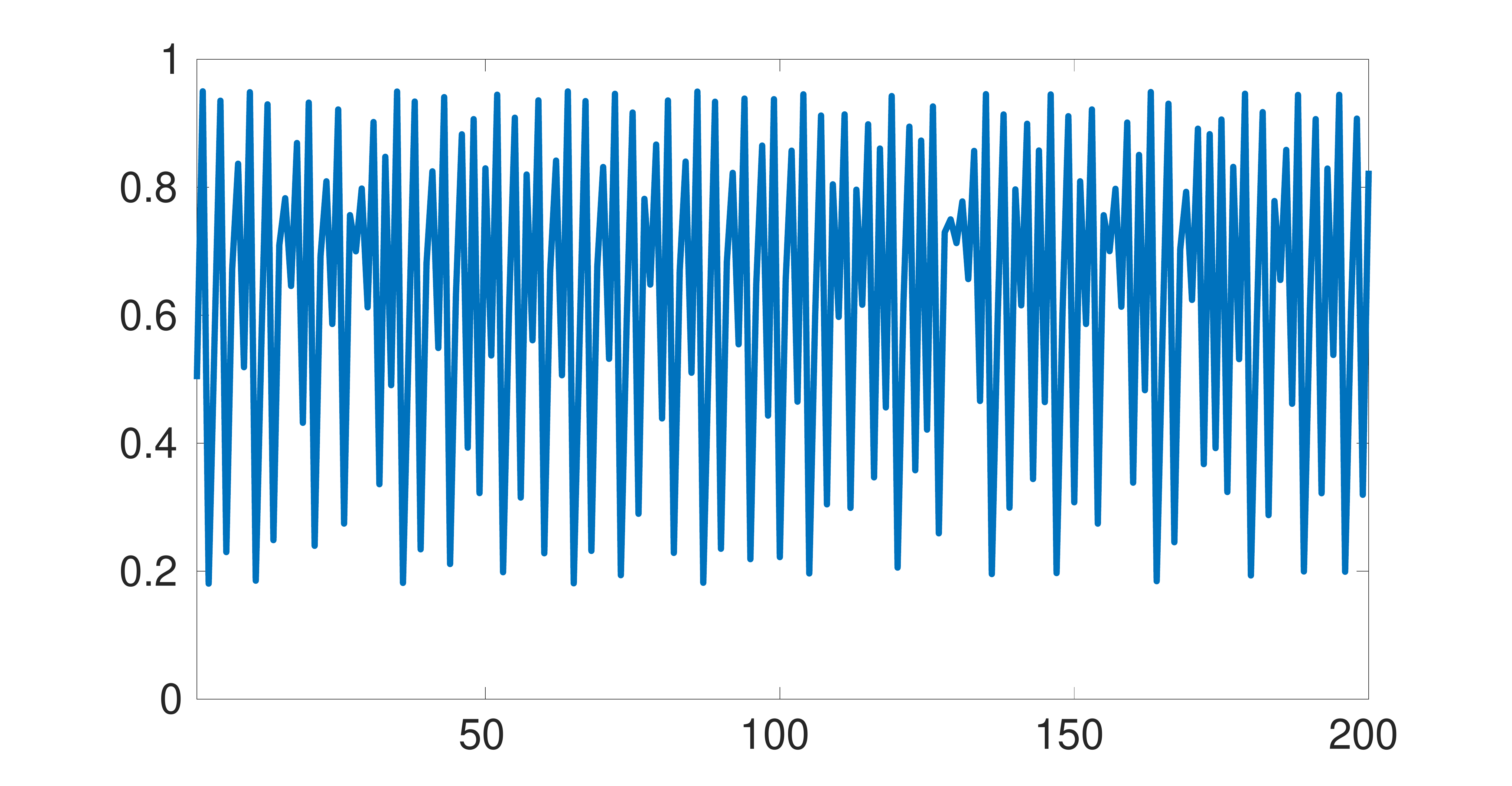}
\put(50,-1){$n$}
\put(1,26){$x_n$}
\end{overpic}
\end{minipage}
\caption{A periodic (top) and a chaotic (bottom) signal of the logistic map of size two hundred with $\mu = 3.5$ and $\mu = 3.8$, respectively.}
\label{fig_discrete_sequence}
\end{figure}

In this section we consider two discrete dynamical systems called the logistic map and the sine-circle map. The first one is the logistic map, popularized by Robert May~\cite{may1976simple}, which is defined by the sequence
\begin{equation} \label{eq_logistic}
x_{n+1}=\mu x_n(1-x_n),\quad x_0 = 0.5,
\end{equation}
where $\mu$ is the bifurcation parameter varying between zero and four. This system exhibits periodic or chaotic behaviour depending on the value of $\mu$. Periodic and chaotic signals of the logistic map are plotted in \cref{fig_discrete_sequence}.

\begin{figure}[htbp]
\centering
\begin{minipage}{0.49\textwidth} 
\begin{overpic}[width=\textwidth]{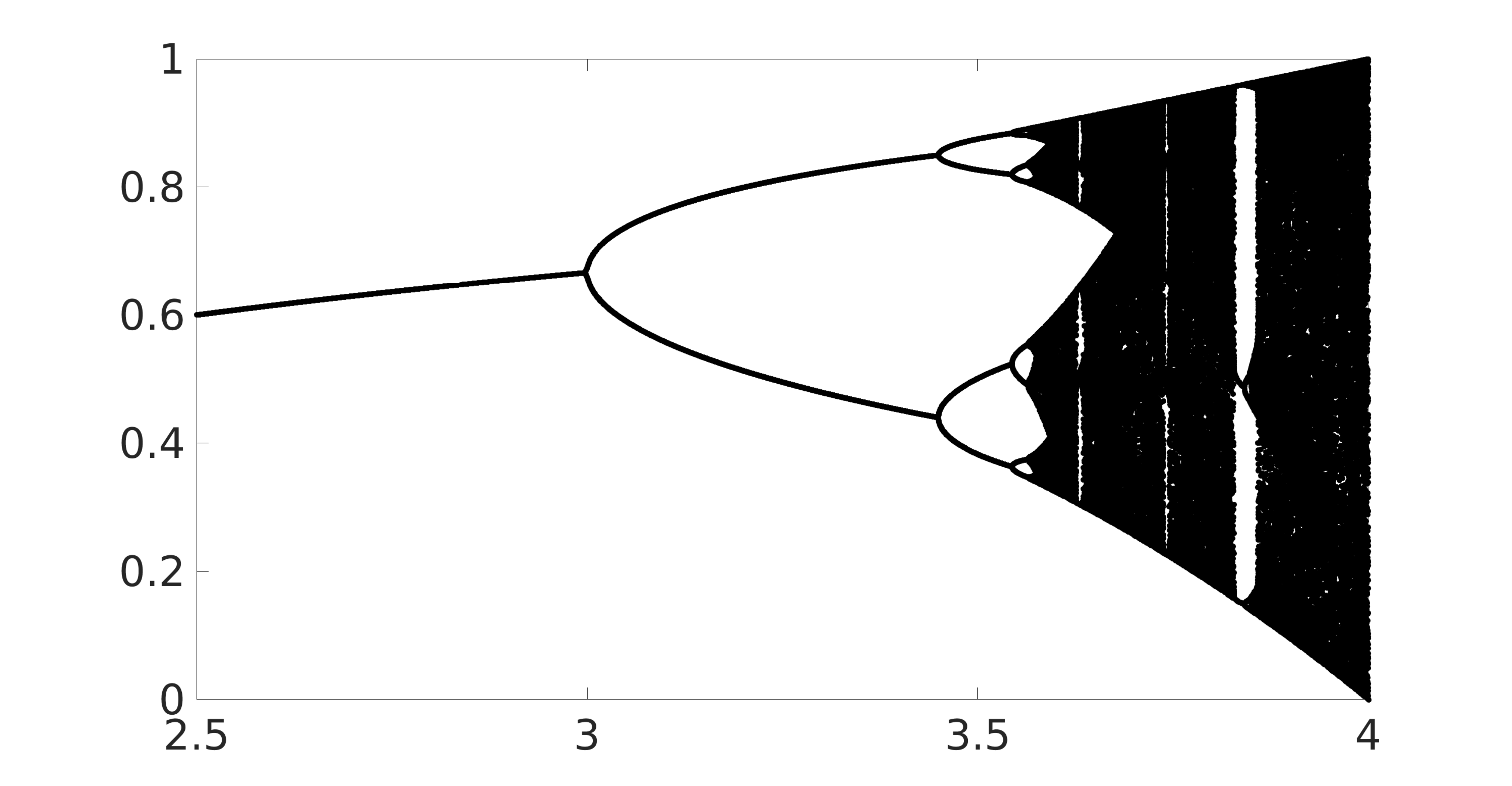}
\put(50,-1){$\mu$}
\put(1,26){$x_n$}
\end{overpic}
\end{minipage}\\
\vspace{0.2cm}
\begin{minipage}{0.49\textwidth} 
\begin{overpic}[width=\textwidth]{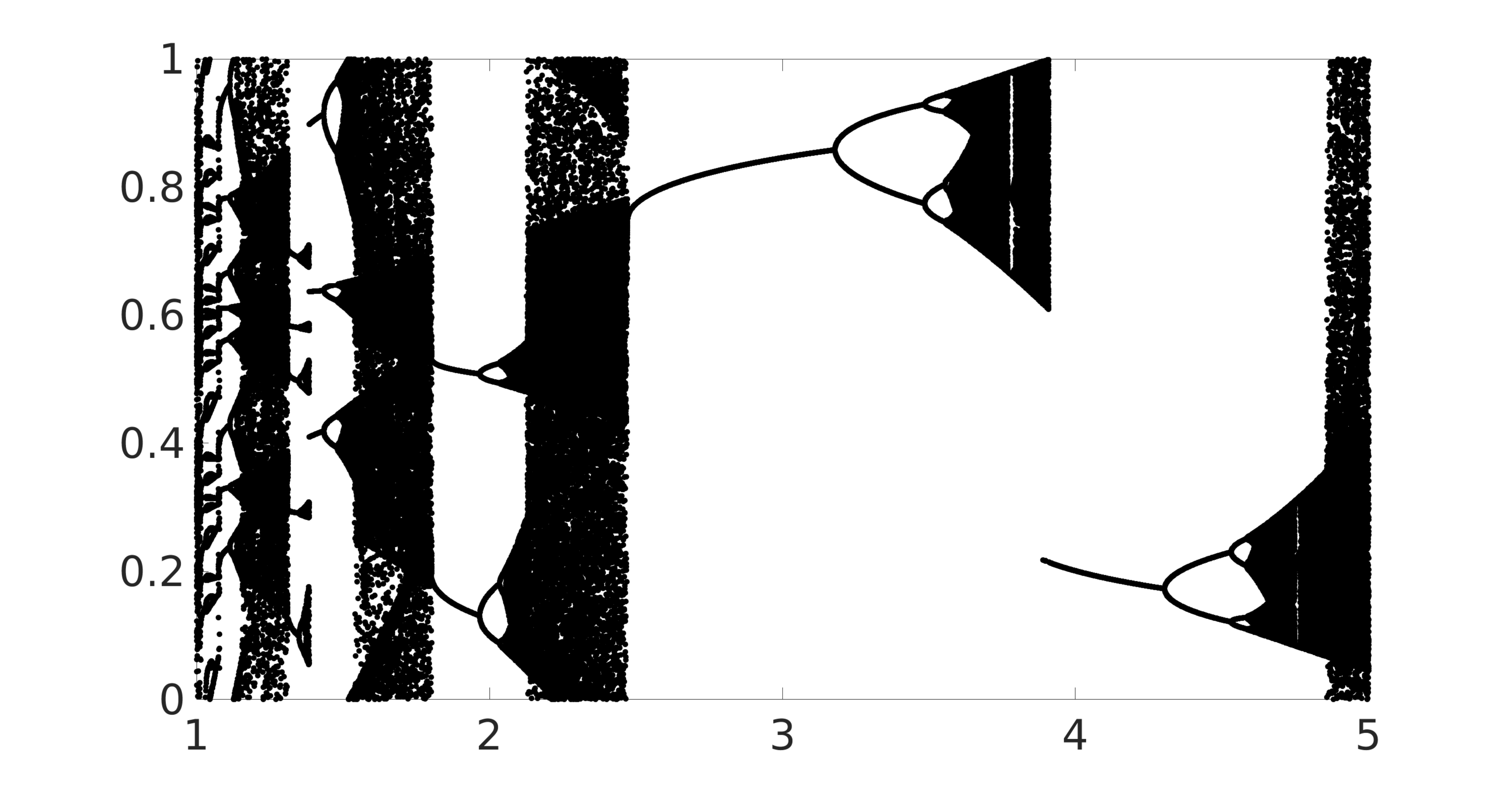}
\put(50,-1){$\mu$}
\put(1,26){$\theta_n$}
\end{overpic}
\end{minipage}
\caption{Bifurcation diagrams of the logistic map (top) and the sine-circle map (bottom).}
\label{fig_discrete_diagram}
\end{figure}

The bifurcation diagram showing the orbits of the logistic map is represented in \cref{fig_discrete_diagram} (top). The behaviour of the attractors for different parameters $\mu$ has been extensively studied~\cite[Chap.~10]{strogatz2014nonlinear} and a highlight is the period-doubling cascade happening for $\mu\in [0,3.54409]$.

\begin{figure}[htbp]
\centering
\begin{minipage}{0.49\textwidth} 
\begin{overpic}[width=\textwidth]{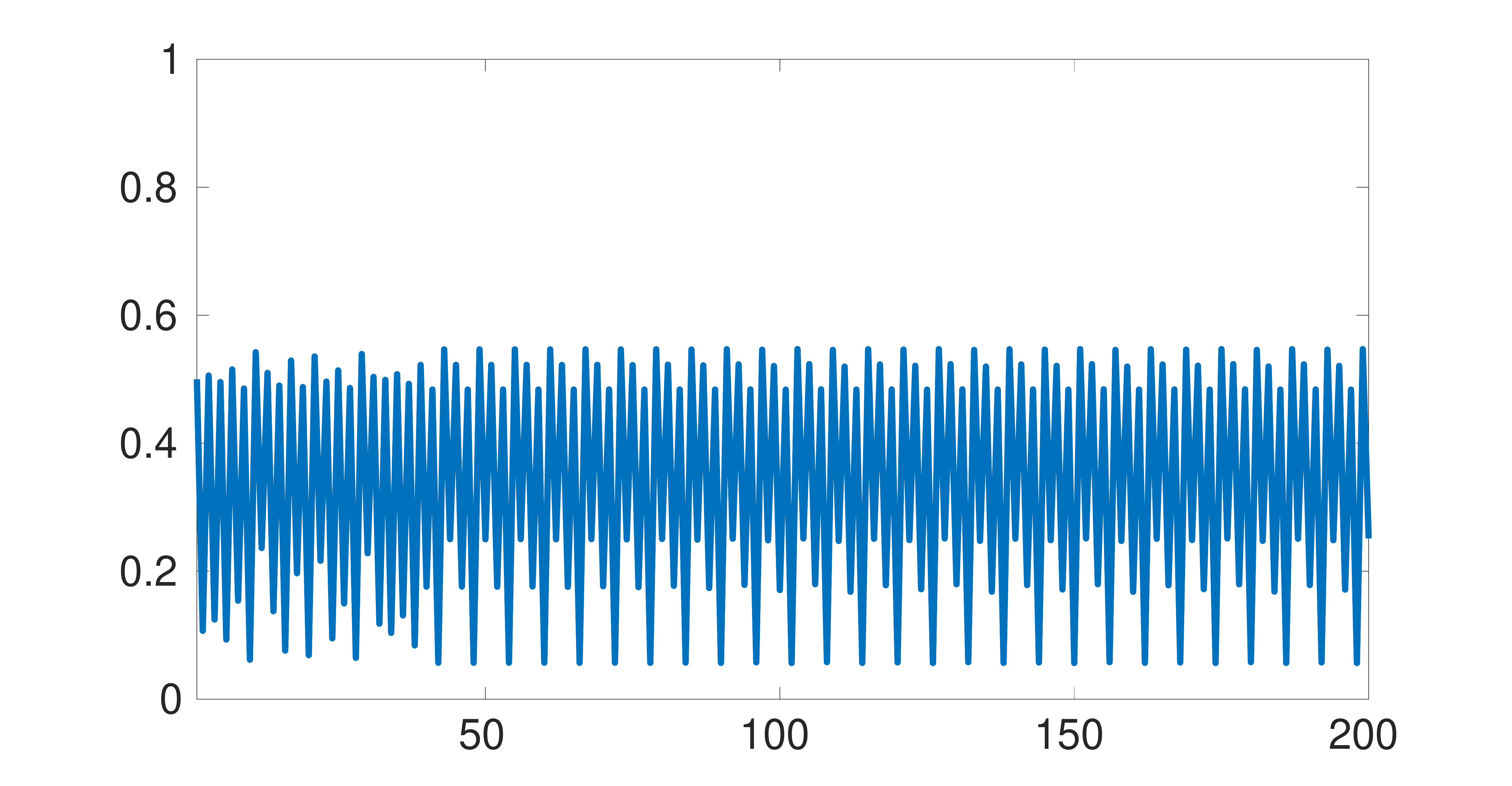}
\put(51,-1){$n$}
\put(1,26){$\theta_n$}
\end{overpic}
\end{minipage}\\
\vspace{0.2cm}
\begin{minipage}{0.49\textwidth} 
\begin{overpic}[width=\textwidth]{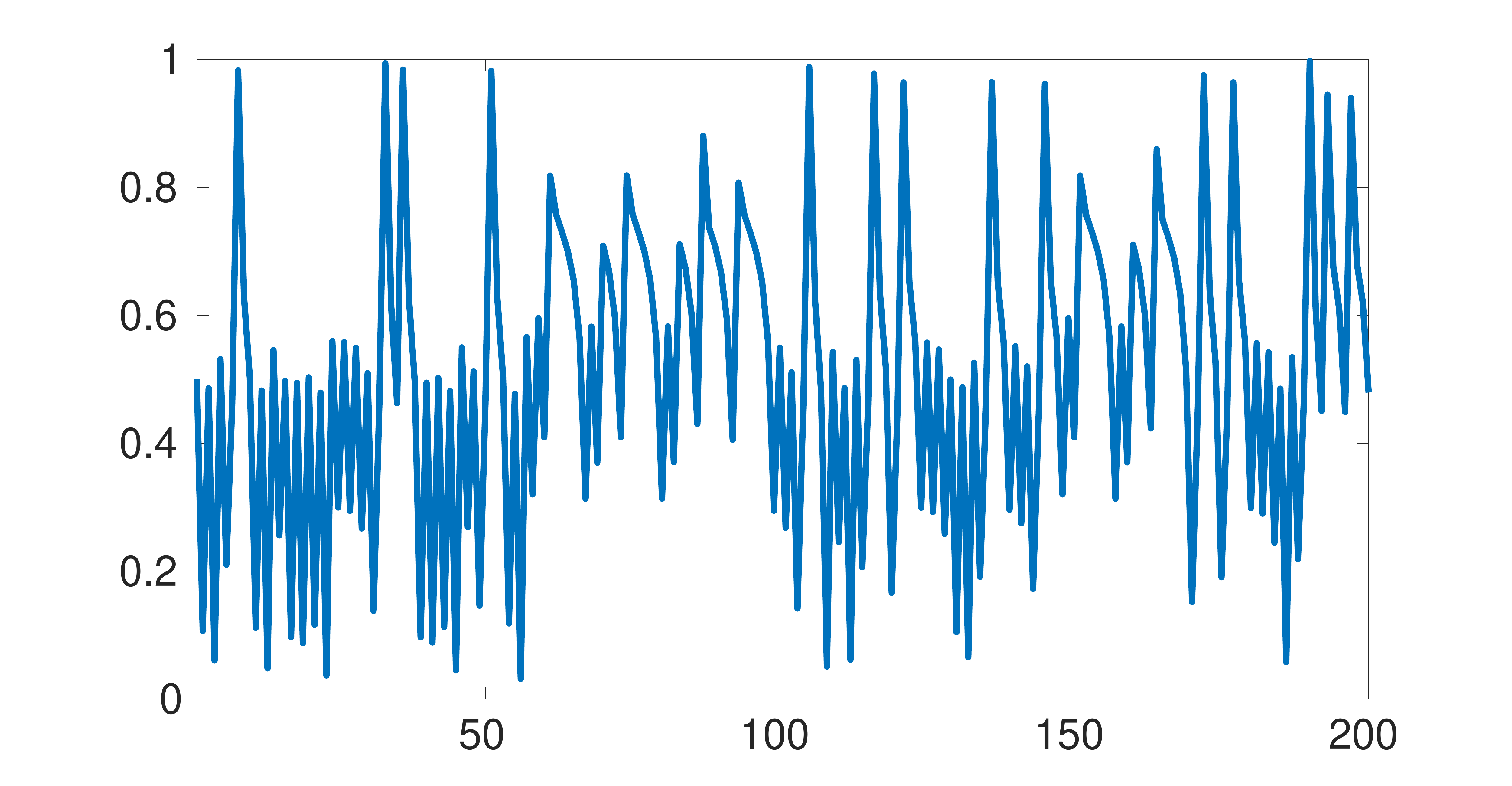}
\put(50,-1){$n$}
\put(1,26){$\theta_n$}
\end{overpic}
\end{minipage}
\caption{A periodic (top) and a chaotic (bottom) signal of the sine-circle map of size two hundred with $\mu = 2.1$ and $\mu = 2.3$, respectively.}
\label{fig_discrete_sequence2}
\end{figure}

The second dynamical system considered in this section is the sine-circle map~\cite[Chap.~6]{hilborn2000chaos}, which is sometimes referred to as the circle map. It takes the form of the following nonlinear map
\begin{equation} \label{eq_sine_circle}
\theta_{n+1}=\theta_n+\Omega-\frac{\mu}{2\pi}\sin(2\pi\theta_n)\mod [1],\quad \theta_0 = 0.5,
\end{equation}
where $\Omega = 0.606661$ and $\mu\in[0,5]$ is the parameter that measures the strength of the nonlinearity. Similarly to the logistic map, iterating \cref{eq_sine_circle} leads to periodic or chaotic signals depending on the bifurcation parameter $\mu$ chosen. \cref{fig_discrete_sequence2} illustrates two signals with different behaviours, generated using a bifurcation parameter of $\mu = 2.1$ (top) and $\mu = 2.3$ (bottom).
The bottom panel of \cref{fig_discrete_diagram} shows the bifurcation diagram of the sine-circle map.

We now want to classify signals generated by the logistic and sine-circle maps according to their chaotic and non-chaotic behaviour. Our main goal, and challenge, is to find a neural network that is able to learn the features characterising chaotic signals of the logistic map and generalise on signals generated by the sine-circle map. To do this, we generate two data sets by computing signals of length one thousand of the logistic and sine-circle maps. This is done for five thousand different values of the parameter $\mu$, uniformly distributed in the interval $[0,5]$. The logistic (resp. sine-circle) dataset is then composed of $24\%$ (resp. $35\%$) of chaotic time series. First, we randomize the logistic map data set across the bifurcation parameter and we choose two thirds of the data to be the training set. Then, with the rest (one third) of the logistic map data set, we test the training of the five neural networks described in \cref{sec_Net_TSC}. Note that to classify the time series of the sine-circle data set we use these five neural networks, which have been trained on the logistic map training set.

\begin{figure}[htbp]
\centering
\begin{minipage}{0.49\textwidth} 
\begin{overpic}[width=\textwidth]{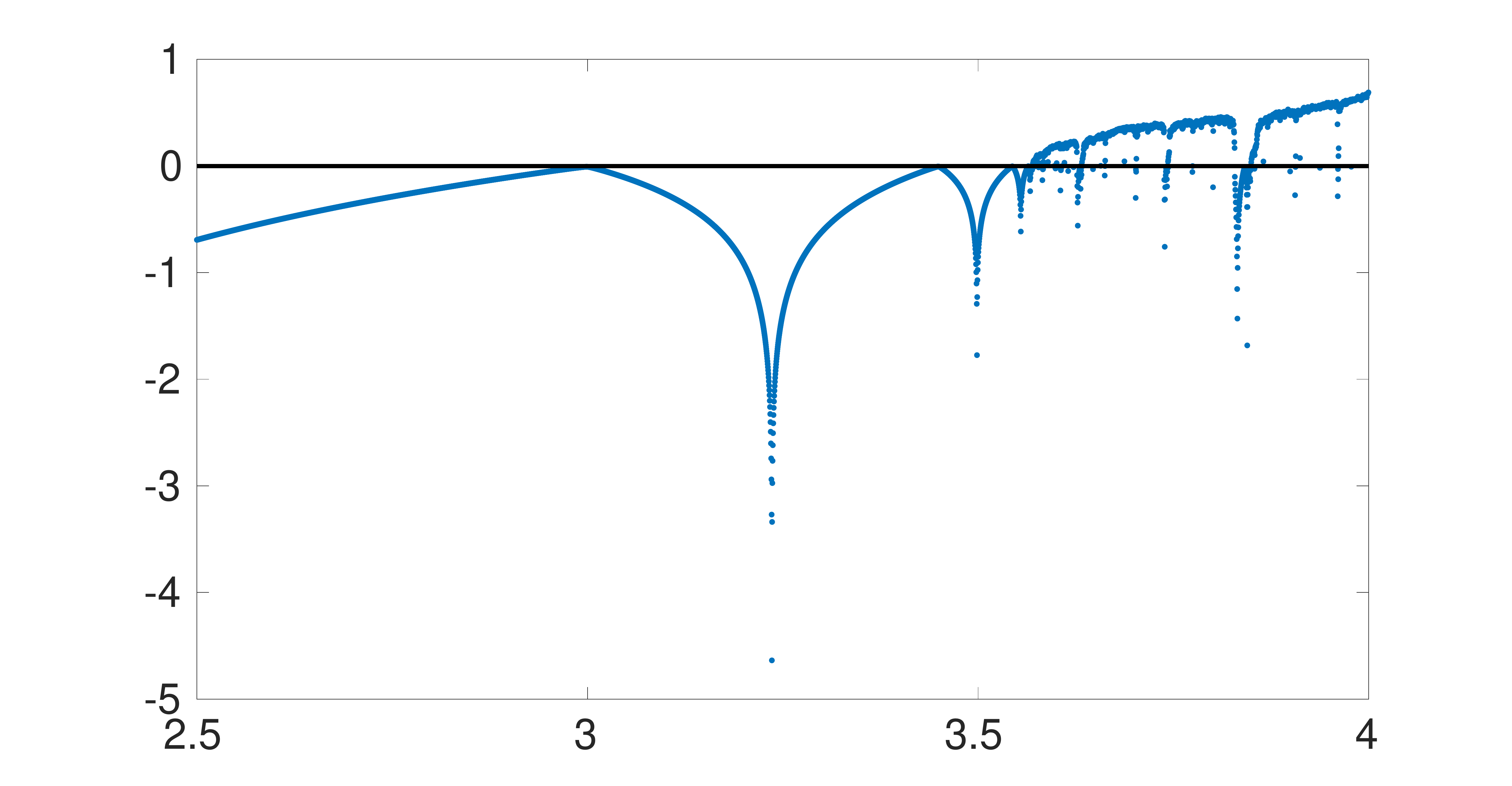}
\put(2,25){$\lambda$}
\put(51,1){$\mu$}
\end{overpic}
\end{minipage}\\
\vspace{0.2cm}
\begin{minipage}{0.49\textwidth} 
\begin{overpic}[width=\textwidth]{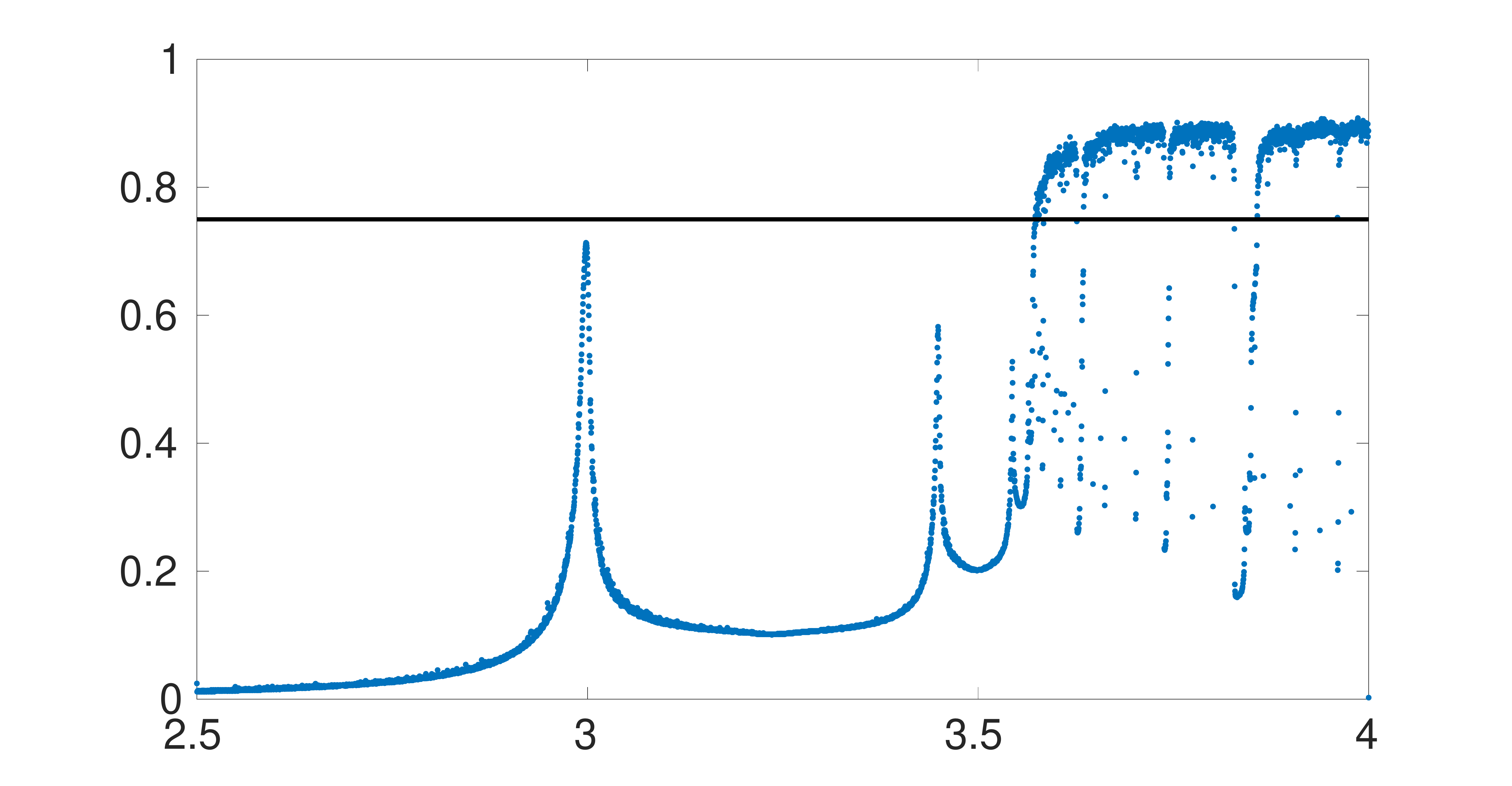}
\put(1,19){\rotatebox{90}{Entropy}}
\put(51,1){$\mu$}
\end{overpic}
\end{minipage}
\caption{Lyapunov exponents (top) and Shannon entropy (bottom) of the logistic map. A time series is chaotic if its Lyapunov exponent is greater than zero and its entropy greater than $0.75$. The horizontal black lines in the plots indicate these thresholds.}
\label{fig_log_lyap}
\end{figure}

The classification of the time series is done using two measures from dynamical systems theory. The first measure is the Lyapunov exponent which is defined as 
\begin{equation}
 \lambda = \lim_{n \to +\infty} \frac{1}{n}\sum_{i=0}^{n-1}\log|f'(x_i)|
\end{equation}
for a discrete dynamical system $x_{n+1} = f(x_n)$ and expresses the exponential separation, viz. $d(t) = d_0e^{\lambda t}$, of two nearby trajectories originally separated by distance $d_0 = \epsilon \ll 1$ at time $t = 0$. The second measure is the Shannon entropy, which uses the probability distribution function of a trajectory to quantify a range of accessible states for a dynamical system and relates to the potential topological transitivity of the system~\cite[Chap.~9]{hilborn2000chaos}. Hence, we expect a chaotic system to have well distributed trajectories in space compared to a periodic one and the aim is then to count the number of accessible states for the system. Thus, we define the Shannon entropy of a time series $x_n$ to be
\begin{equation}
 S_N = - \frac{1}{\log(N)}\sum_r p_r \log(p_r),
\end{equation}
where $p_r$ is the probability to be on the state $r$ reached by the system with $p_r = \frac{1}{N}\#\{x_i = r | 1 \leq i \leq N\}$ and $N$ is the number of sample points. Here, the entropy $S_N$ has been normalized to lie in $[0,1]$ so that the entropy of a constant signal $S_N \to 0$ while the entropy of a chaotic time series $S_N \to 1$.

We classify a given signal as chaotic when its Lyapunov exponent is strictly positive and its entropy is greater than a given threshold, experimentally set at $0.75$ (see also \cref{fig_log_lyap}), and non-chaotic otherwise. It is crucial to classify the training data set accurately in order to reduce misclassifications on the testing set. On that note, by using the Shannon entropy in addition to the Lyapunov exponent as a measure of chaos we gained an incremental improvement in accuracy. This is because the Lyapunov exponent was misclassifying some quasi-periodic signals as chaotic.

The Lyapunov exponent and the Shannon entropy of the logistic map as a function of the bifurcation parameter $\mu$ are illustrated in \cref{fig_log_lyap}. In real applications, computing these quantities over the whole range of parameters and in some cases without knowing the expression of the underlying dynamical system can be unfeasible or computationally expensive, which justifies the approach of using a machine learning algorithm to perform the classification automatically.

\begin{table*}[htbp]
\centering
\caption{Classification accuracy on the logistic and sine-circle maps data sets. The neural networks are trained on logistic signals and the accuracy is averaged over five training cycles.}
\label{tab_discrete}
\vspace{0.2cm}
\begin{tabular}{lccccc}
\hline
Networks    & ShallowNet & MLP  &  FCN & ResNet & LKCNN \\
\hline
Logistic    &       99.5 & 83.4 & 95.3 & {96.7} &    {98.8} \\
Sine-circle &       64.9 & 60.2 & 54.0 & {44.8} &    {89.8} \\
\hline
\end{tabular}
\end{table*}

The average classification accuracy of the neural networks ShallowNet, MLP, FCN, ResNet, and LKCNN is reported in \cref{tab_discrete}. The ShallowNet, MLP, FCN, and ResNet architectures classify signals from the sine-circle map with an accuracy less than $65\%$. The LKCNN network however seems to override overfitting issues on the training set by capturing the main features of chaotic and periodic signals and gets an average classification accuracy of $89.8\%$. It is of interest to notice that the shallow neural network reaches an accuracy greater than state-of-the-art time series classification networks on the sine-circle data set despite its simplicity. Improving the accuracy of LKCNN on the sine-circle map might be challenging because this dynamical system leads to signals with behaviour that is absent in the training set of the logistic map (see e.g. the regime $\mu\in [1,1.3]$ in \cref{fig_discrete_diagram} (bottom)).

\section{Continuous dynamical systems} \label{sec_continuous_res}

We now consider continuous dynamical systems of ordinary and partial differential equations that exhibit temporal and spatiotemporal chaos, respectively. The aim here is to determine whether a neural network trained on a low dimensional dynamical system is able to generalise and classify univariate time series generated by a higher dimensional dynamical system. We will first consider the Lorenz system since it is one of the most typical continuous dynamical systems with a chaotic behaviour which has been widely studied in the twentieth century~\cite{sparrow2012lorenz}.

\subsection{Lorenz system} \label{sec_lorenz}

The Lorenz system~\cite{lorenz1963deterministic} consists of the following three ordinary differential equations:
\begin{subequations}
\label{eq_lorenz}
\begin{align}
\dot{x} &= \sigma(y-x),\\
\dot{y} &= x(\rho-z)-y,\\
\dot{z} &=xy-\beta z.
\end{align}
\end{subequations}
Taking the parameters $\sigma = 10$, $\beta = 8/3$, and varying $\rho$ in $[0,250]$ yields convergent, periodic, and chaotic solutions. We numerically solve \cref{eq_lorenz} using MATLAB's function \texttt{ode45} with $[x,y,z] = [1,1,1]$ as initial condition. Integrating the equations for $t\in[0,100]$ we obtain time series for $x(t)$, $y(t)$, and $z(t)$ of length one thousand, and we carry out this operation for five thousand values of the bifurcation parameter $\rho$ in the range $[0,250]$.

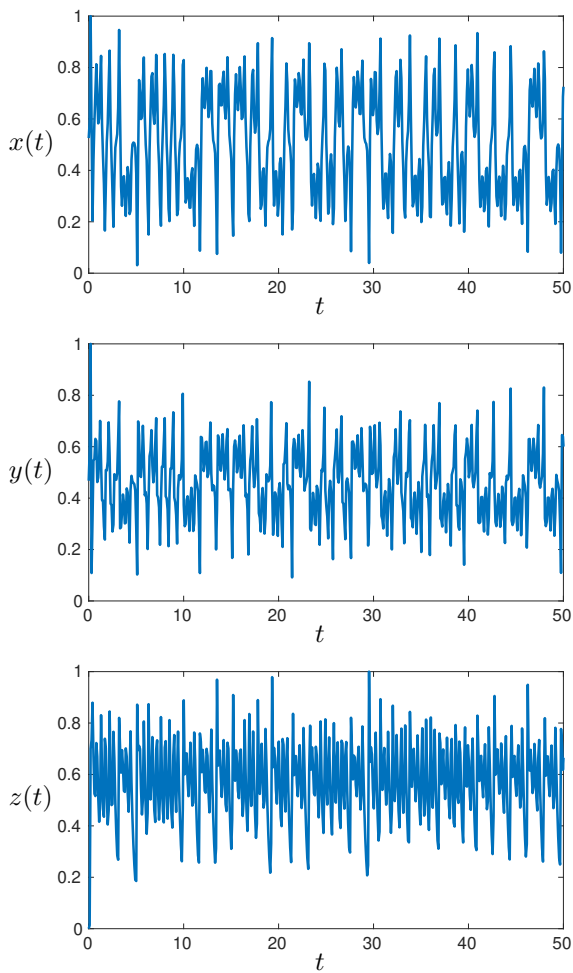
\begin{figure}[htbp]
\centering
\begin{minipage}{0.49\textwidth} 
\begin{overpic}[width=\textwidth]{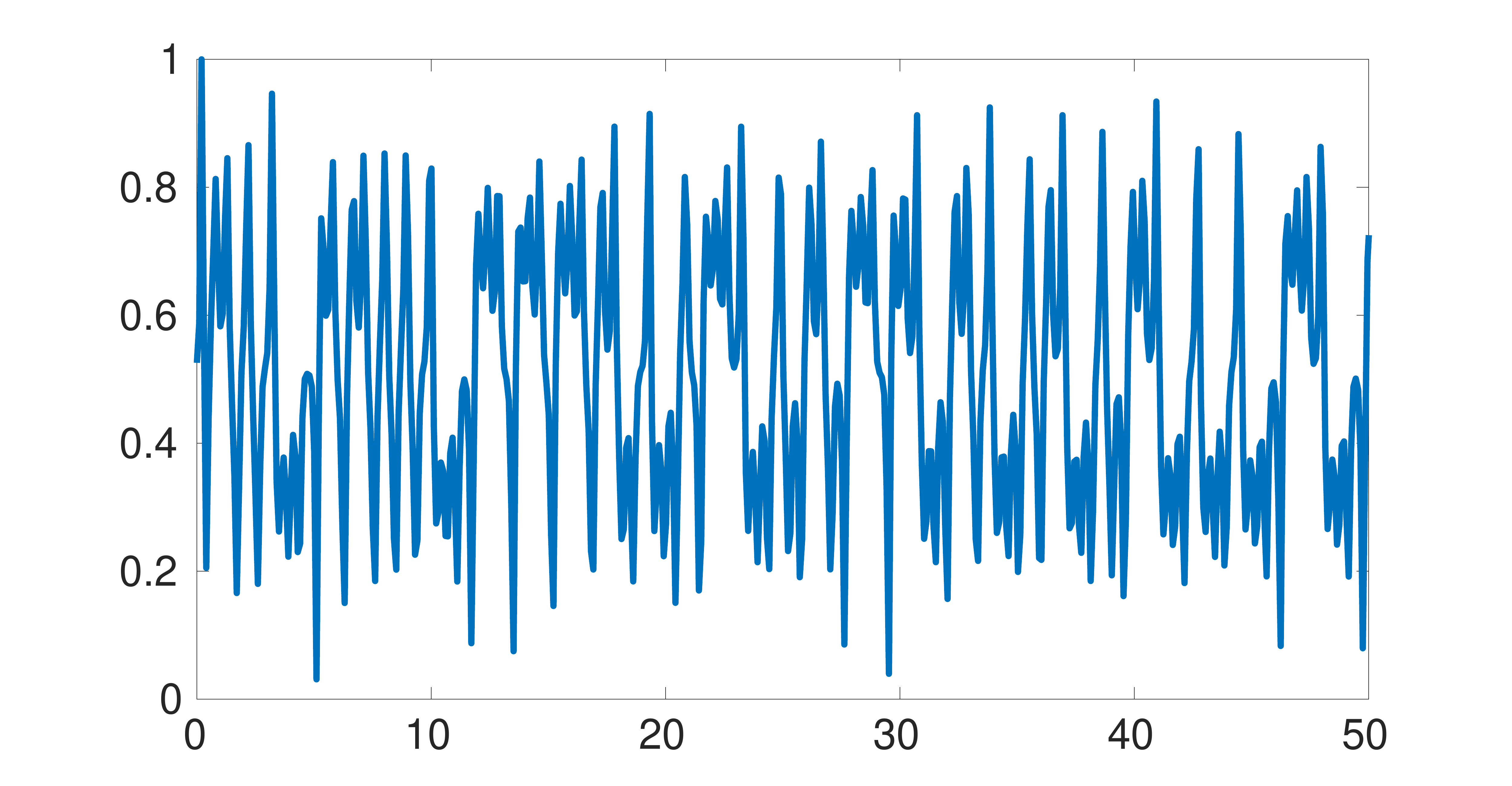}
\put(50,-1){$t$}
\put(0,26){$x(t)$}
\end{overpic}
\end{minipage}\\
\vspace{0.2cm}
\begin{minipage}{0.49\textwidth} 
\begin{overpic}[width=\textwidth]{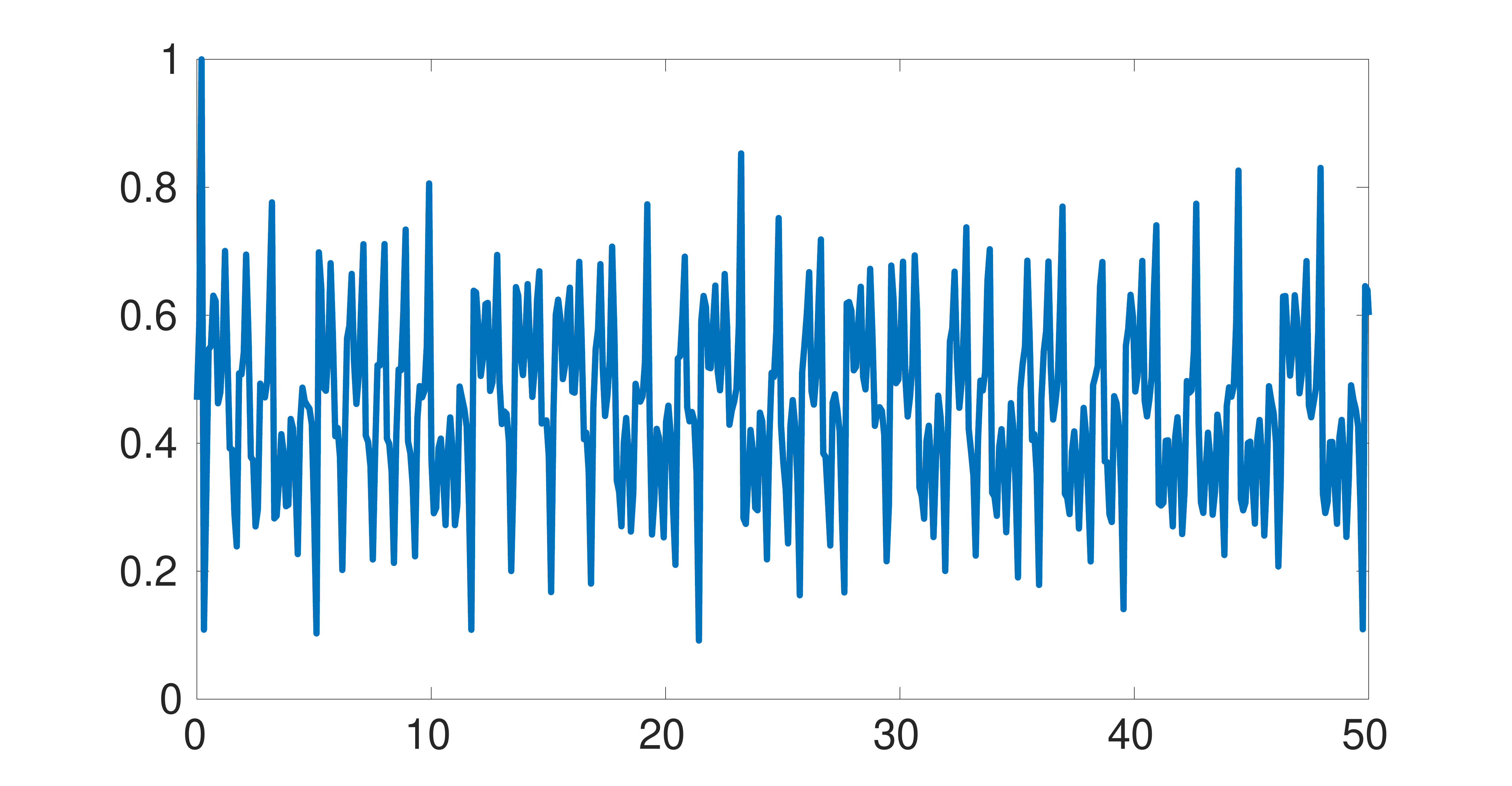}
\put(50,-1){$t$}
\put(0,26){$y(t)$}
\end{overpic}
\end{minipage}\\
\vspace{0.2cm}
\begin{minipage}{0.49\textwidth} 
\begin{overpic}[width=\textwidth]{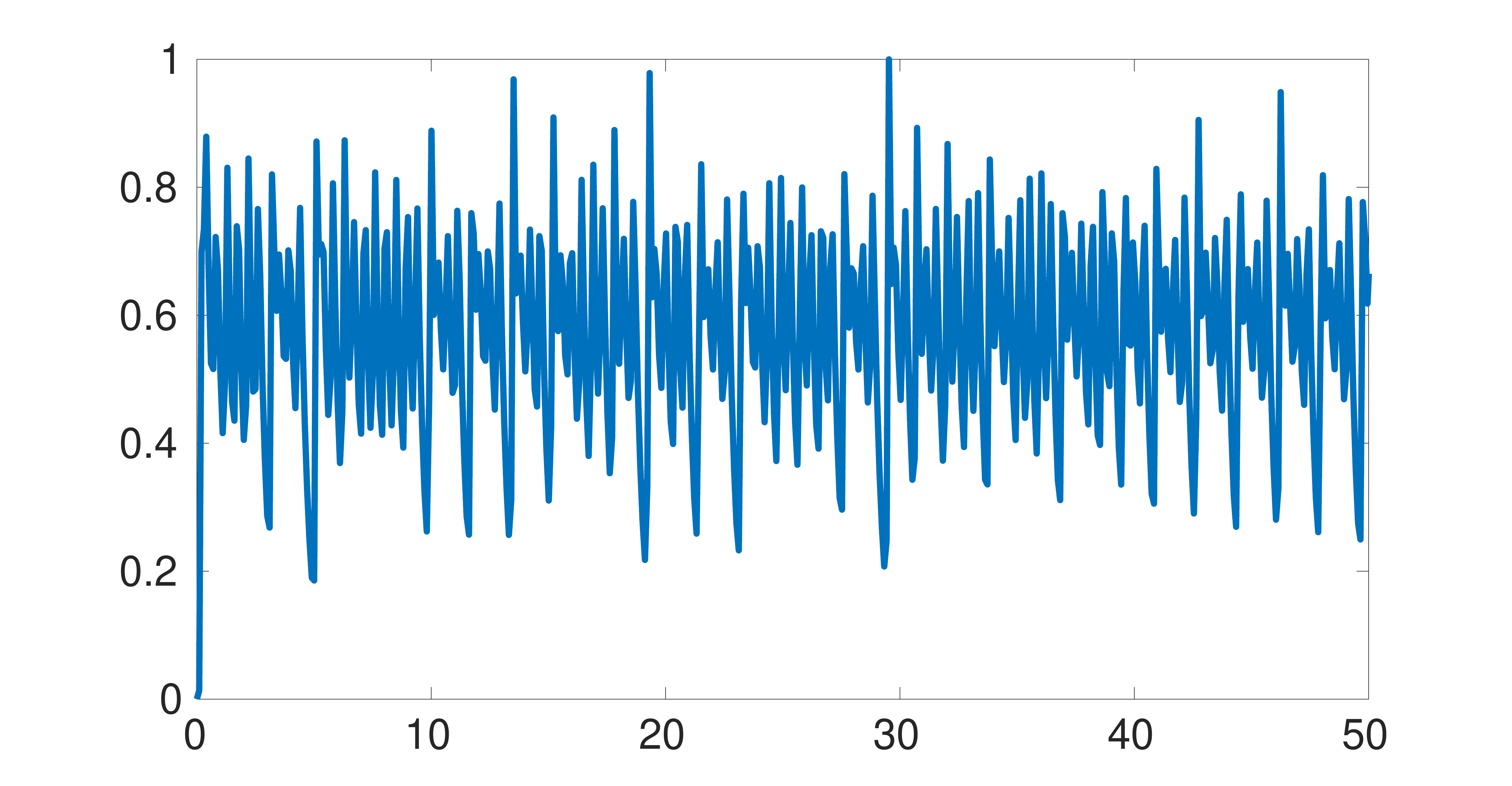}
\put(50,-1){$t$}
\put(0,26){$z(t)$}
\end{overpic}
\end{minipage}
\caption{Normalized time series of the $x$, $y$, and $z$ components of the Lorenz system with bifurcation parameter $\rho = 70$.}
\label{fig_lorenz_xyz}
\end{figure}

\begin{figure*}[htbp]
\centering
\begin{minipage}{0.49\textwidth} 
\begin{overpic}[width=\textwidth]{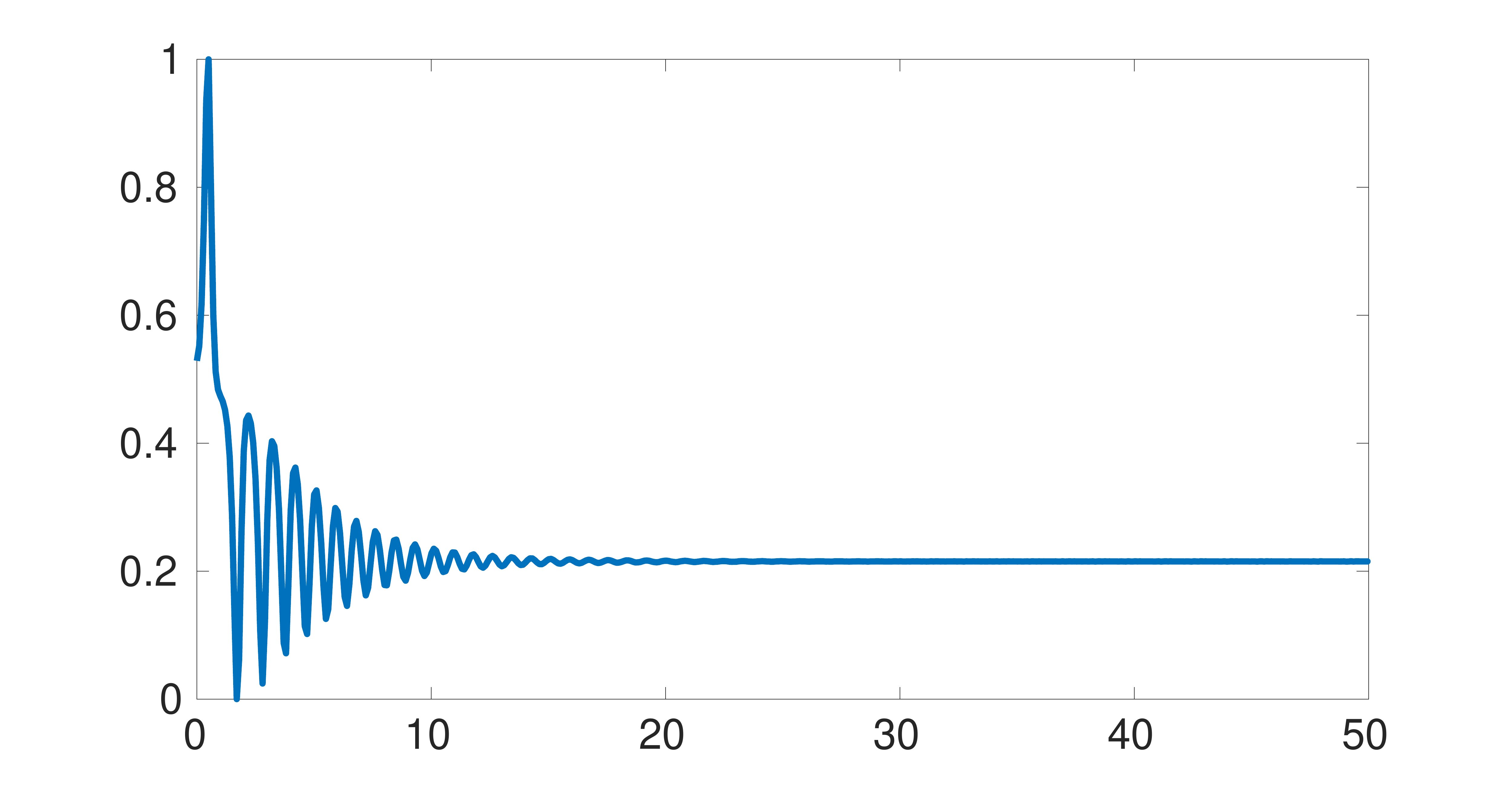}
\put(51,-2){$t$}
\put(-2,25){$x(t)$}
\put(0,48){(a)}
\end{overpic}
\end{minipage}
\begin{minipage}{0.49\textwidth} 
\begin{overpic}[width=\textwidth]{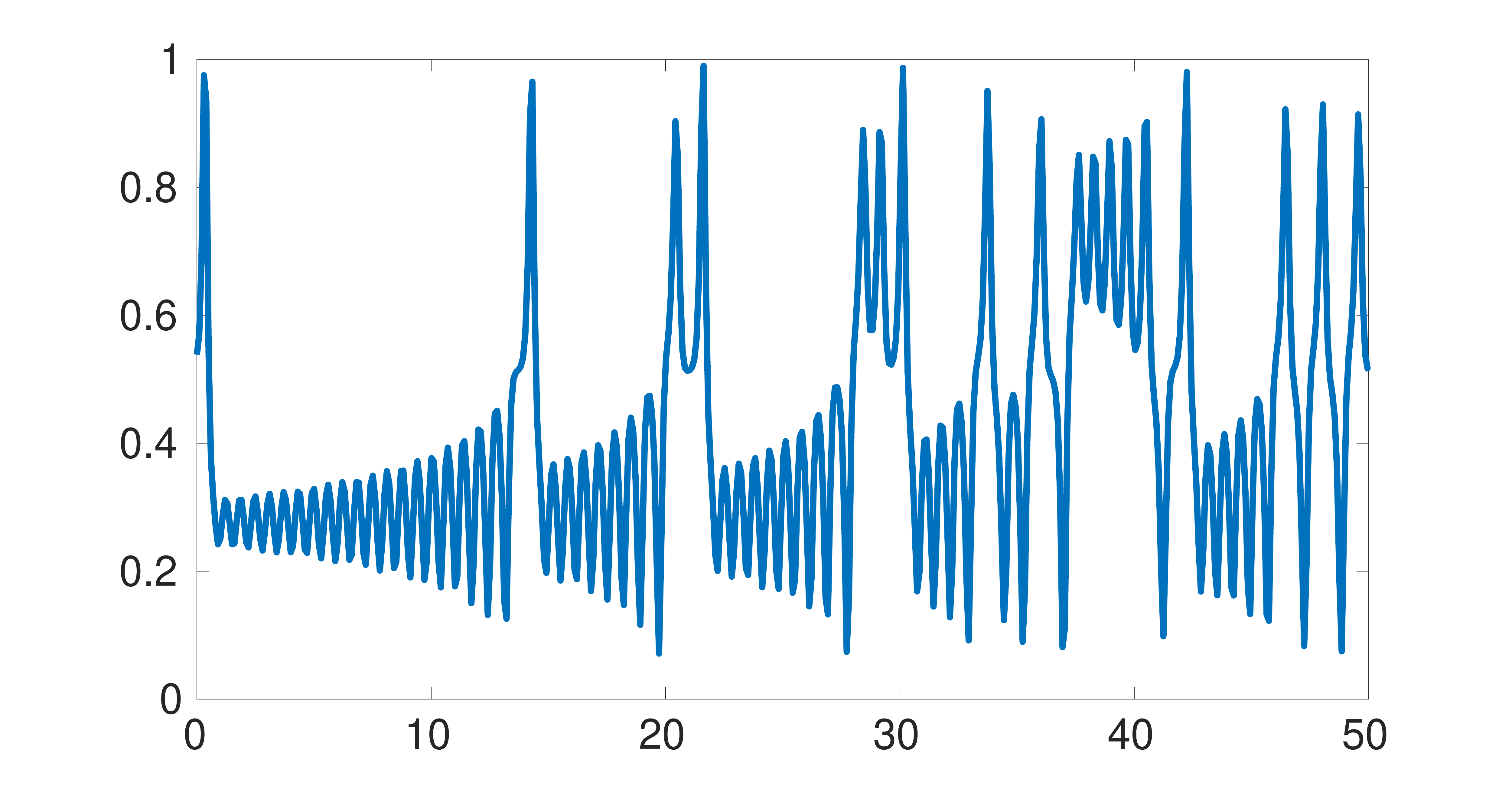}
\put(51,-2){$t$}
\put(-2,25){$x(t)$}
\put(0,48){(b)}
\end{overpic}
\end{minipage}\\
\vspace{0.2cm}
\begin{minipage}{0.49\textwidth} 
\begin{overpic}[width=\textwidth]{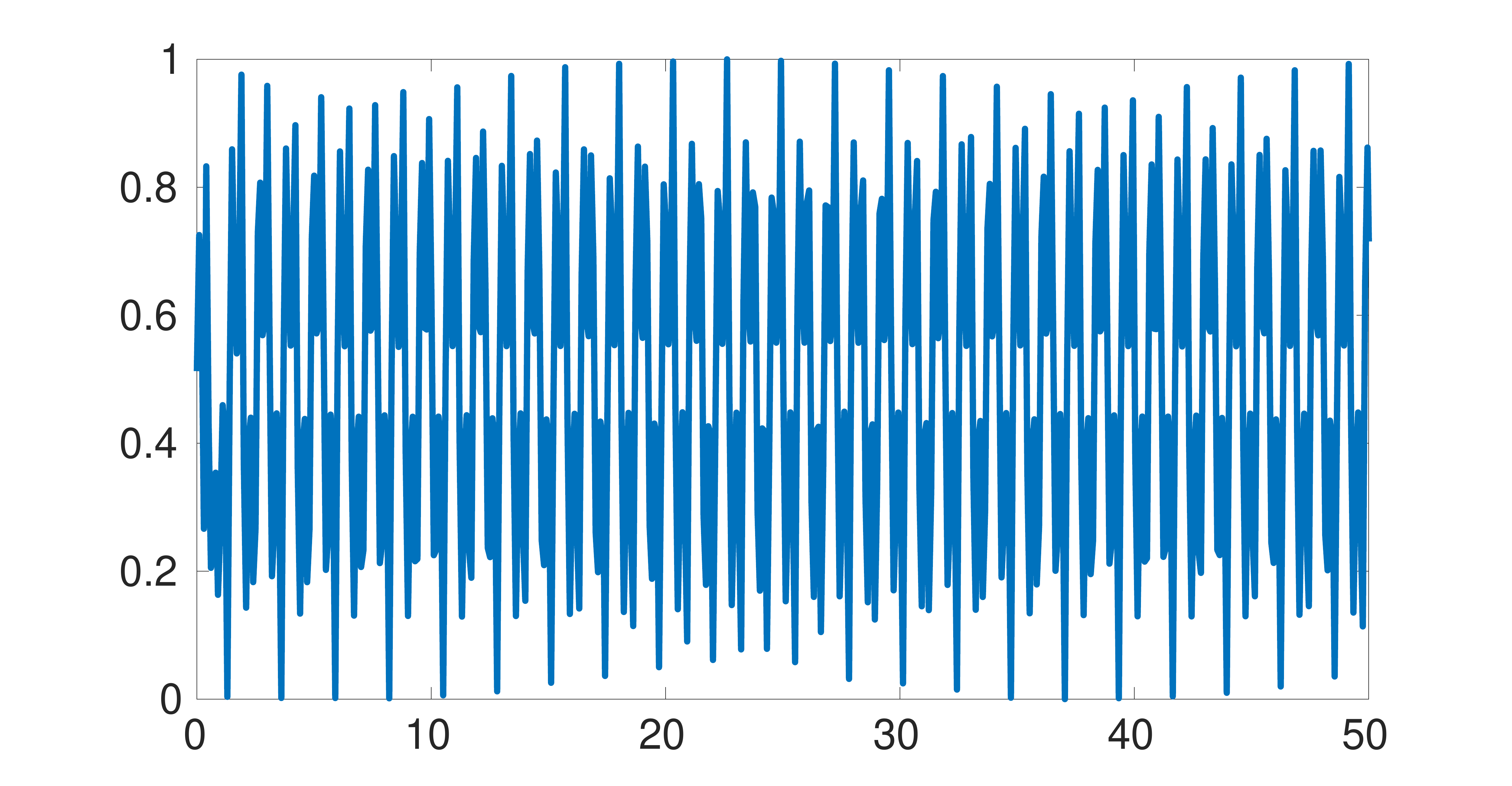}
\put(51,-2){$t$}
\put(-2,25){$x(t)$}
\put(0,48){(c)}
\end{overpic}
\end{minipage}
\begin{minipage}{0.49\textwidth} 
\begin{overpic}[width=\textwidth]{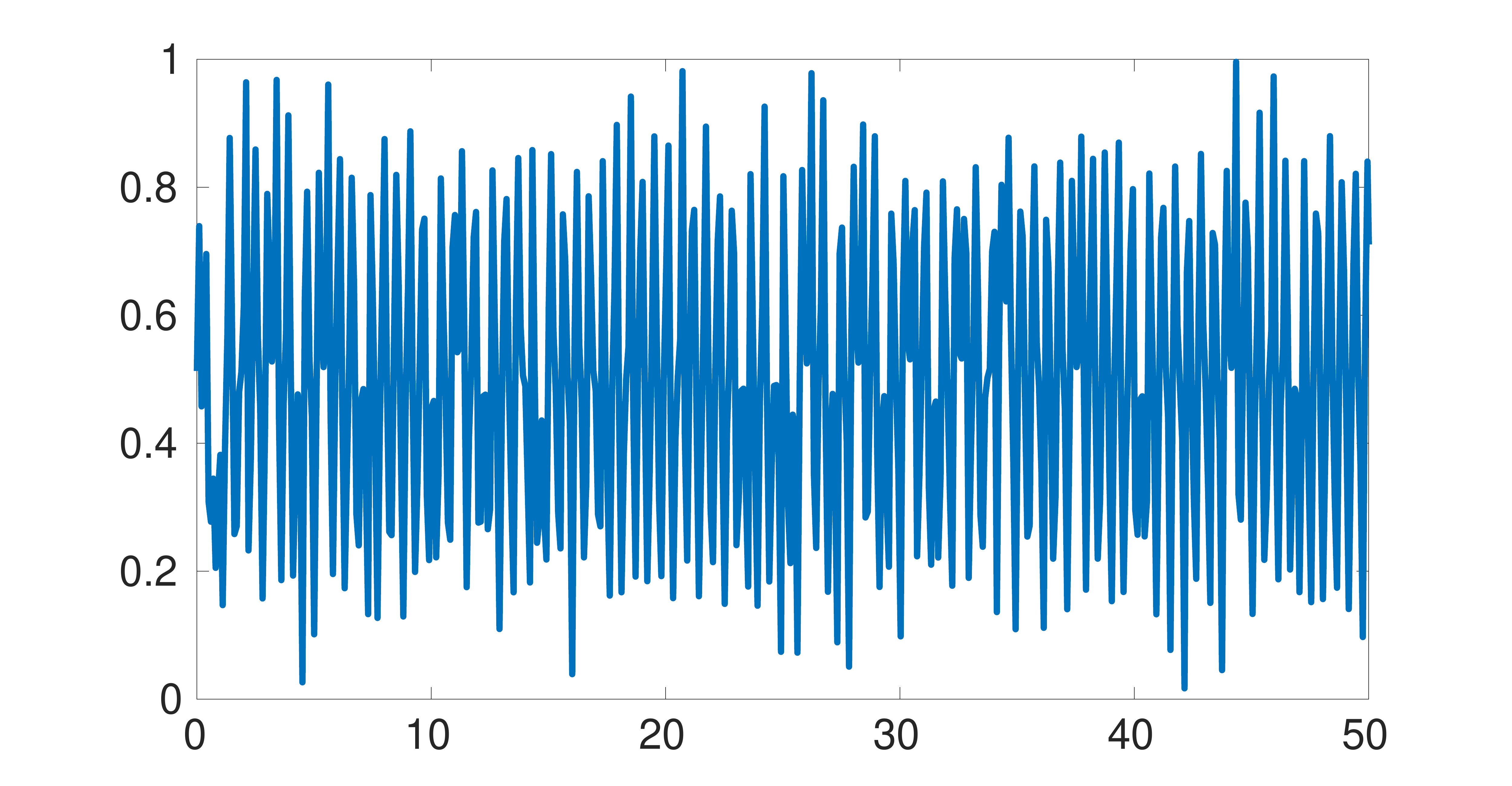}
\put(51,-2){$t$}
\put(-2,25){$x(t)$}
\put(0,48){(d)}
\end{overpic}
\end{minipage}
\caption{Normalized time series of the $x$ component of the Lorenz system with bifurcation parameter $\rho = 15$ (a), $28$ (b), $160$ (c), and $180$ (d).}
\label{fig_lorenz_sequence}
\end{figure*}
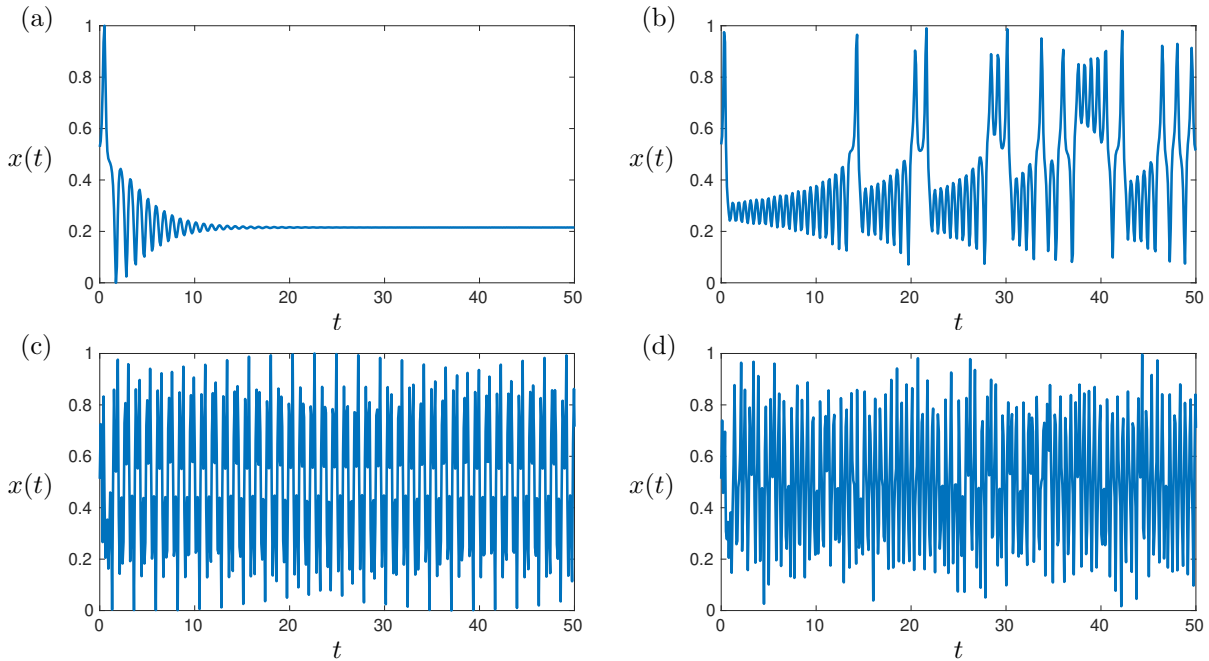

The time series $x(t)$, $y(t)$, and $z(t)$ are normalized by the linear transformation $x(t)\mapsto (x(t)-m)/(M-m)$, where $M$ and $m$ are respectively the maximum and minimum of the time series, such that their range are in the interval $[0,1]$ (see time series in \cref{fig_lorenz_xyz} for $\rho = 70$). Note that normalising the time series is crucial to obtain good generalisation performance because the $x$, $y$, and $z$ time series do not have the same range of values. \cref{fig_lorenz_sequence} depicts four time series of the variable $x(t)$ generated by numerically solving \cref{eq_lorenz} for $\rho = 15$, $28$, $160$, and $180$.

\begin{figure}[htbp]
\centering
\begin{overpic}[width=0.49\textwidth]{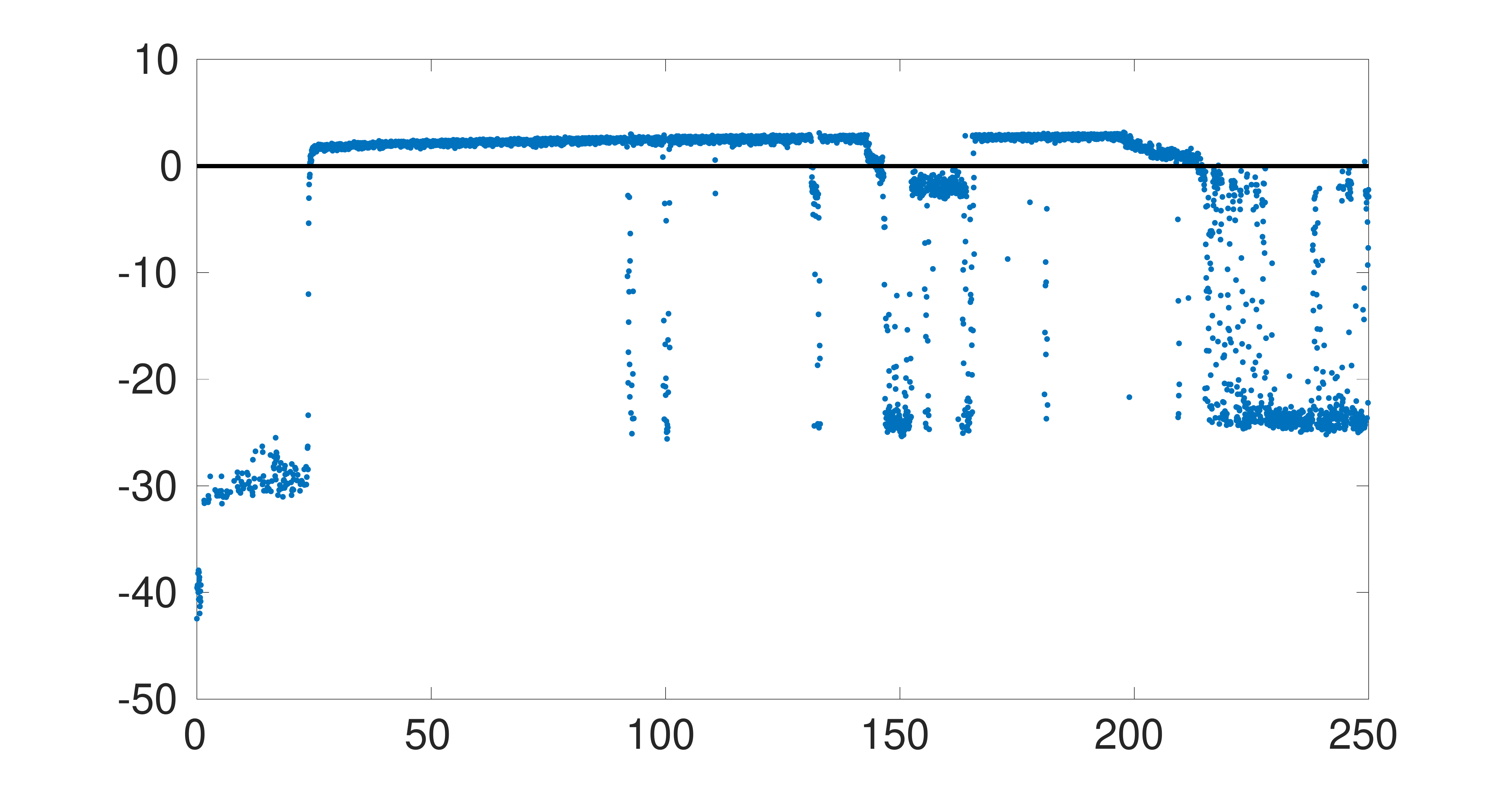}
\put(4,25){$\lambda$}
\put(51,1){$\rho$}
\end{overpic}
\caption{Lyapunov exponents of the $x(t)$ component of the Lorenz system for $\sigma = 10$, $\beta = 8/3$, and $\rho\in [0,250]$. A positive Lyapunov exponent (points above the horizontal black line) indicates a chaotic solution to the Lorenz equations.}
\label{fig_lorenz_lyap}
\end{figure}

We classify the time series of the Lorenz system as chaotic or non-chaotic according to the sign of the Lyapunov exponent at the corresponding regimes of the bifurcation parameter $\rho$ in order to generate training and testing data sets for the neural networks. Here, we compute the Lyapunov exponents for the time series of the variable $x(t)$ starting from some initial condition $x(0)$ as follows
\begin{equation}
 \lambda = \lim_{t \to +\infty}\lim_{\epsilon \to 0} \frac{1}{t}\log\lt(\frac{|x(t) - x_\epsilon(t)|}{\epsilon} \rt),
\end{equation}
where $|x(0) - x_\epsilon(0)| < \epsilon \ll 1$. \cref{fig_lorenz_lyap} shows the Lyapunov exponents of the variable $x(t)$, which determine the classification of the testing set of time series given to the neural networks described in \cref{sec_Net_TSC}. For example, the chaotic time series plotted in \cref{fig_lorenz_sequence} (d) corresponds to a bifurcation parameter of $\rho = 180$ and has a strictly positive Lyapunov exponent as shown in \cref{fig_lorenz_lyap}. For continuous dynamical systems the Shannon entropy did not appear to be a precise measure of chaotic behaviour. In fact, Shannon entropy requires a threshold to determine if a given time series is chaotic or not. Setting a good threshold might not always be possible and requires a lot of experimentation. Moreover, in this case, the distribution of values of the time series is continuous, giving a broad probability distribution function, which makes the task of choosing a threshold troublesome. Therefore, we do not consider it to classify the time series of the Lorenz system.

The different neural networks are trained on time series of the $x$ component of the Lorenz system and tested on the $y$ and $z$ components. These three datasets are composed of two thirds of chaotic time series. Similarly to the logistic map data set (see \cref{sec_discrete_res}), the $x$ component set is divided in the following way: two thirds for training and one third for testing. We then compare the classification accuracy of the networks described in \cref{sec_Net_TSC} on the two data sets. The results are presented in \cref{tab_continuous}.

The convolutional neural network LKCNN (see \cref{sec_convnet}) outperforms the other networks on all the testing sets composed by time series of the $x$, $y$, and $z$ components of the Lorenz system. In particular, it is able to generalise well on the $z$ component by determining whether a given time series is chaotic or not correctly with an average accuracy of {$78.2\%$}. The other neural networks seem to overfit the training set and fail to classify time series of the $z$ component correctly. Note that the $y$ component of the Lorenz system is highly correlated with the $x$ component, unlike the $z$ component (see \cref{fig_lorenz_xyz}), which explains the relative good classification accuracy (around $75\%$) of all the neural networks on the $y$ component.

\begin{table*}[htbp]
\centering
\caption{Classification accuracy on the Lorenz system. The network is trained on $x$ component of the Lorenz system and the accuracy is averaged over five training cycles.}
\label{tab_continuous}
\vspace{0.2cm}
\begin{tabular}{lccccc}
\hline
Networks & ShallowNet &  MLP &  FCN & ResNet & LKCNN \\
\hline
Lorenz X &       98.5 & 90.2 & 80.3 &   {88.8} &    {98.6} \\
Lorenz Y &       75.9 & 75.5 & 74.6 &   {73.6} &    {95.4} \\
Lorenz Z &       58.2 & 54.9 & 65.5 &   {45.8} &    {78.2} \\
\hline
\end{tabular}
\end{table*}

\subsection{Kuramoto--Sivashinsky equation}

In this section, we consider the Kuramoto--Sivashinsky (KS) equation, which is an example of a fourth-order nonlinear partial differential equation, which exhibits spatiotemporal chaos. This equation was originally derived by Kuramoto~\cite{kuramoto1978diffusion,kuramoto1975formation,kuramoto1976persistent} and Sivashinsky~\cite{sivashinsky1977nonlinear,sivashinsky1980flame,sivashinsky1983instabilities} to model instabilities in laminar flame fronts and arises in a wide range of physical problems such as plasma physics~\cite{kuramoto1976persistent}, flame propagation~\cite{sivashinsky1977nonlinear}, or free surface film flows~\cite{benney1966long,hooper1985nonlinear,sivashinsky1980irregular}. In particular, we study the Kuramoto--Sivashinsky system normalized to the interval $[0,2\pi]$:
\begin{equation} \label{eq_KS_norm}
\begin{gathered}
u_t+ 4 u_{xxxx}+\alpha\left[u_{xx}+\frac{1}{2}(u_x)^2\right] = 0,\\
u(x,0) = u_0(x),\quad u(x+2\pi,t) = u(x,t),
\end{gathered}
\end{equation}
where $x\in[0,2\pi]$, $t\in\mathbb{R}^+$, and $\alpha$ is the bifurcation parameter.

We refer to the study of the attractors by Hyman and Nicolaenko~\cite{hyman1986kuramoto} and follow the approach of Papageorgiou and Smyrlis~\cite{papageorgiou1991route,smyrlis1991predicting} by considering the initial condition
$u_0(x) = -\sin(x)$ to ensure that the integral of the solution over the spatial domain vanishes. Varying the bifurcation parameter $\alpha$ in \cref{eq_KS_norm} yields a wide range of attracting solutions such as periodic, bimodal, travelling wave, or chaotic, numerically studied in~\cite{hyman1986kuramoto}.

We spatially discretise \cref{eq_KS_norm} using the Fourier spectral method with the $2/3$ dealiasing rule~\cite{orszag1971elimination} and temporally using the ETDRK4 scheme of Cox and Matthews \cite{cox2002exponential}. We use the stiff partial differential equation integrator~\cite{montanelli2016solving} in the Chebfun software~\cite{driscoll2014chebfun} with a spectral resolution of $512$ and a time step of $2.5\times 10^{-4}$ to numerically solve \cref{eq_KS_norm} for $t\in[0,10]$. The regimes we considered are listed below based on the values of the bifurcation parameter $\alpha$:
\begin{enumerate}
\item One hundred values of $\alpha$ are uniformly distributed in each of the following intervals: $[18,22]$, $[23,33]$, $[43,45]$, $[56,65]$, $[95,115]$. These intervals are chosen to cover a wide range of behaviours according to~\cite{hyman1986kuramoto}.
\item Five hundred values of $\alpha$ are uniformly distributed in $[120,130]$.
\end{enumerate}
This leads to a data set of one thousand realisations, equally divided between chaotic and non-chaotic behaviour.

\begin{figure*}[htbp]
\centering
\begin{minipage}{0.49\textwidth} 
\begin{overpic}[width=\textwidth,trim={40 0 0 40},clip]{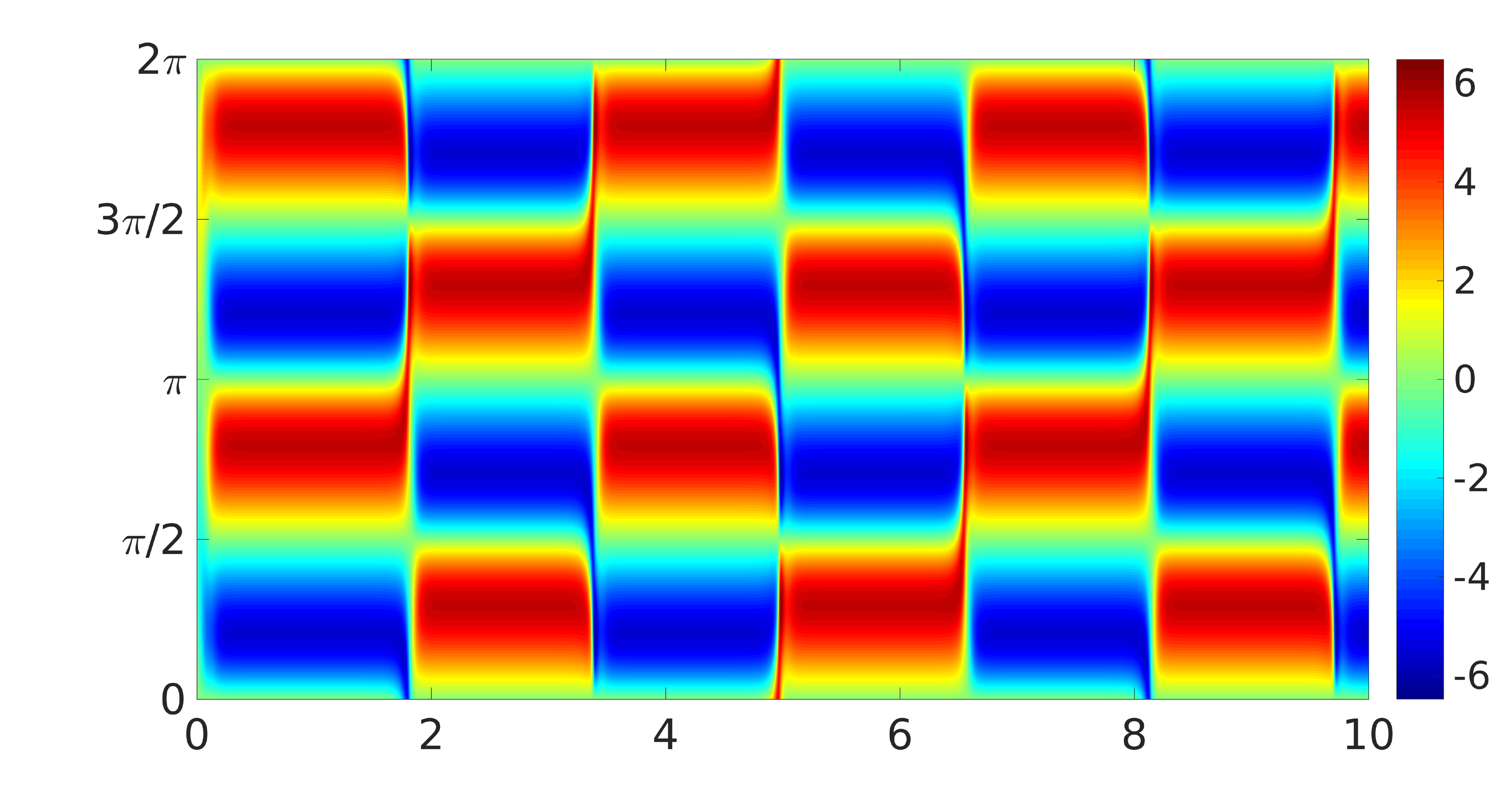}
\put(50,0){$t$}
\put(1,26){$x$}
\put(-2,50){(a)}
\end{overpic}
\end{minipage}
\begin{minipage}{0.49\textwidth} 
\begin{overpic}[width=\textwidth,trim={0 0 40 40},clip]{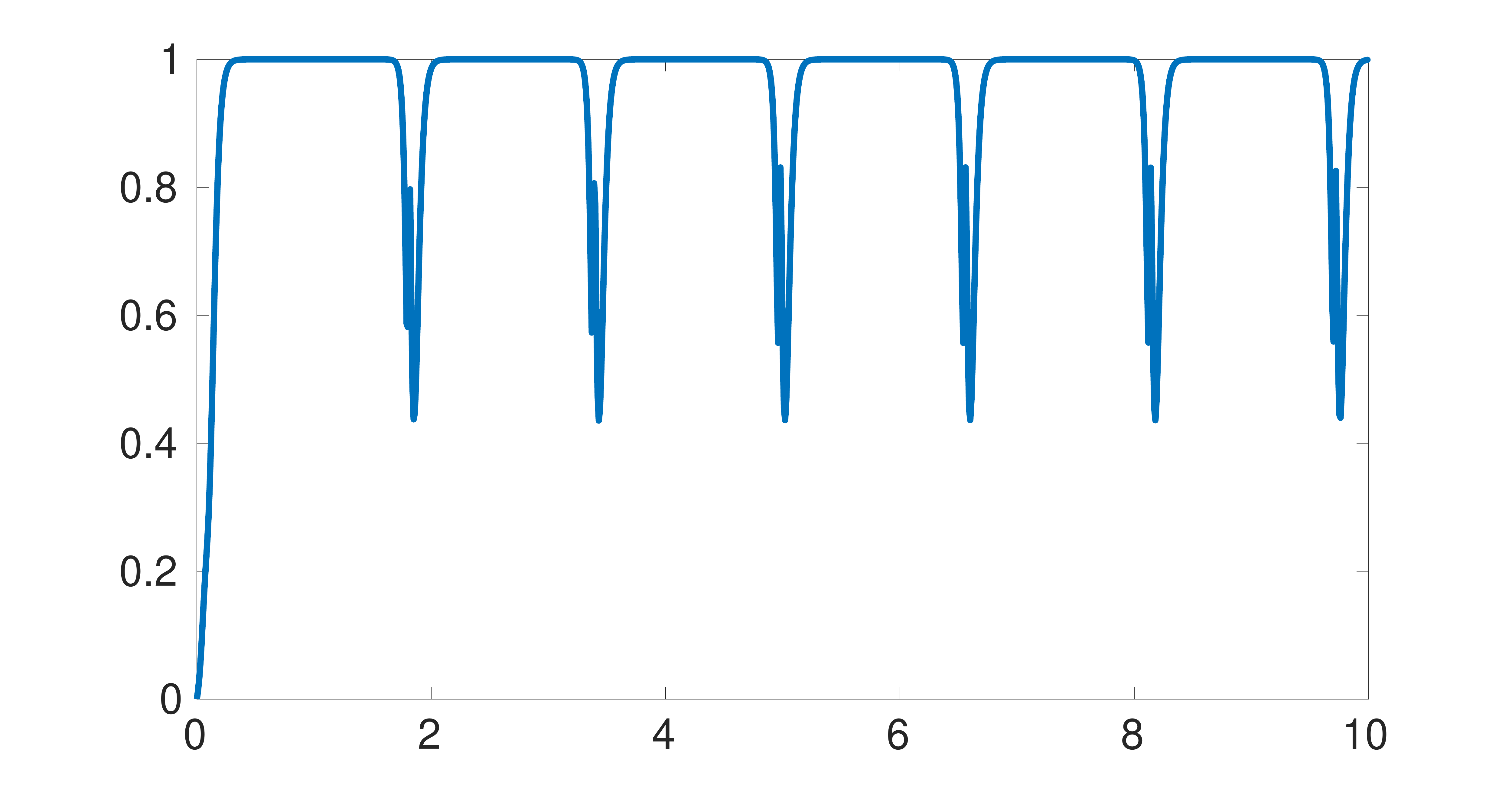}
\put(50,0){$t$}
\put(-1,26){$\mathcal{E}(t)$}
\put(0,50){(e)}
\end{overpic}
\end{minipage}\\
\vspace{0.2cm}
\begin{minipage}{0.49\textwidth} 
\begin{overpic}[width=\textwidth,trim={40 0 0 40},clip]{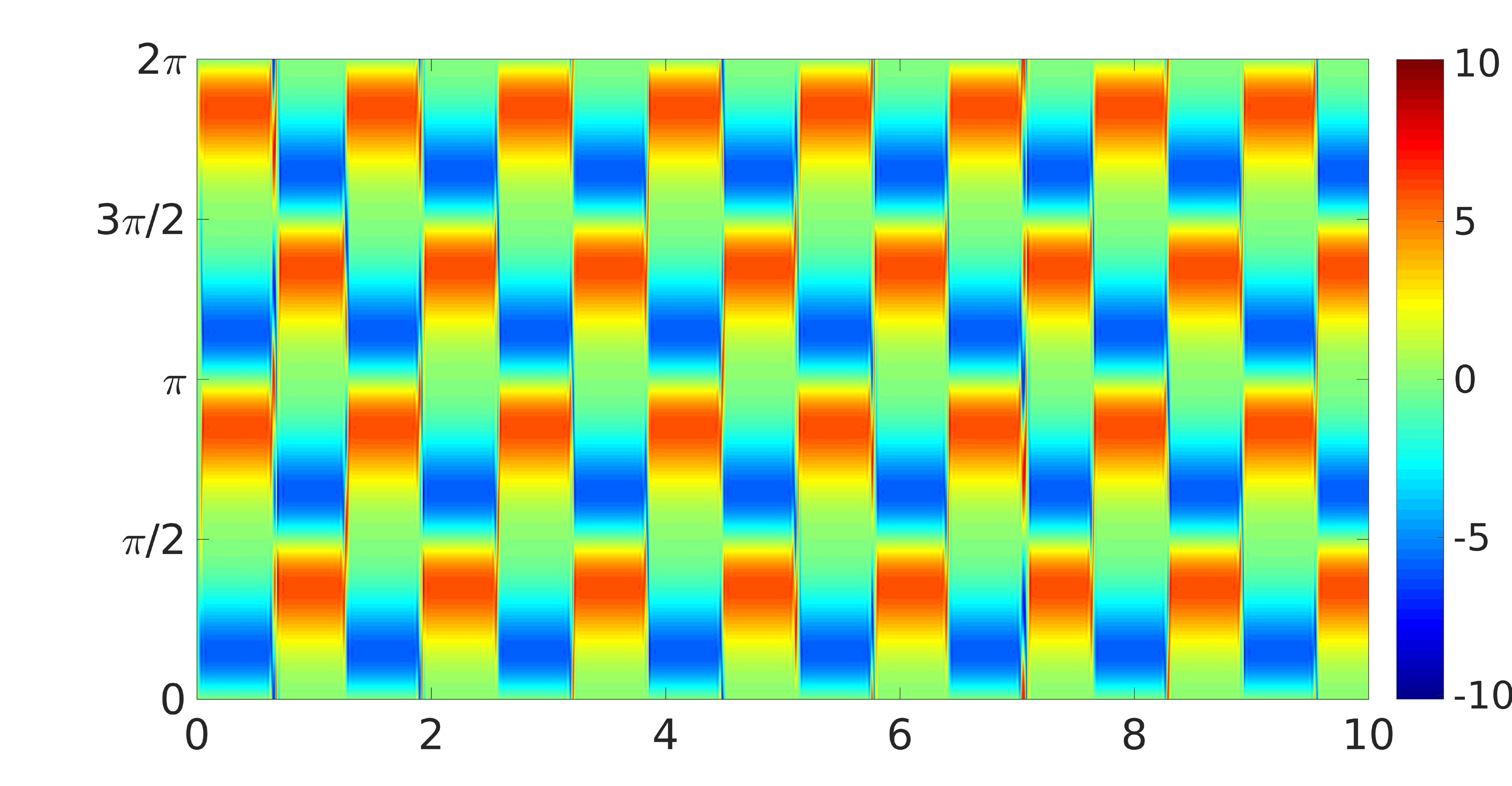}
\put(50,0){$t$}
\put(1,26){$x$}
\put(-2,50){(b)}
\end{overpic}
\end{minipage}
\begin{minipage}{0.49\textwidth} 
\begin{overpic}[width=\textwidth,trim={0 0 40 40},clip]{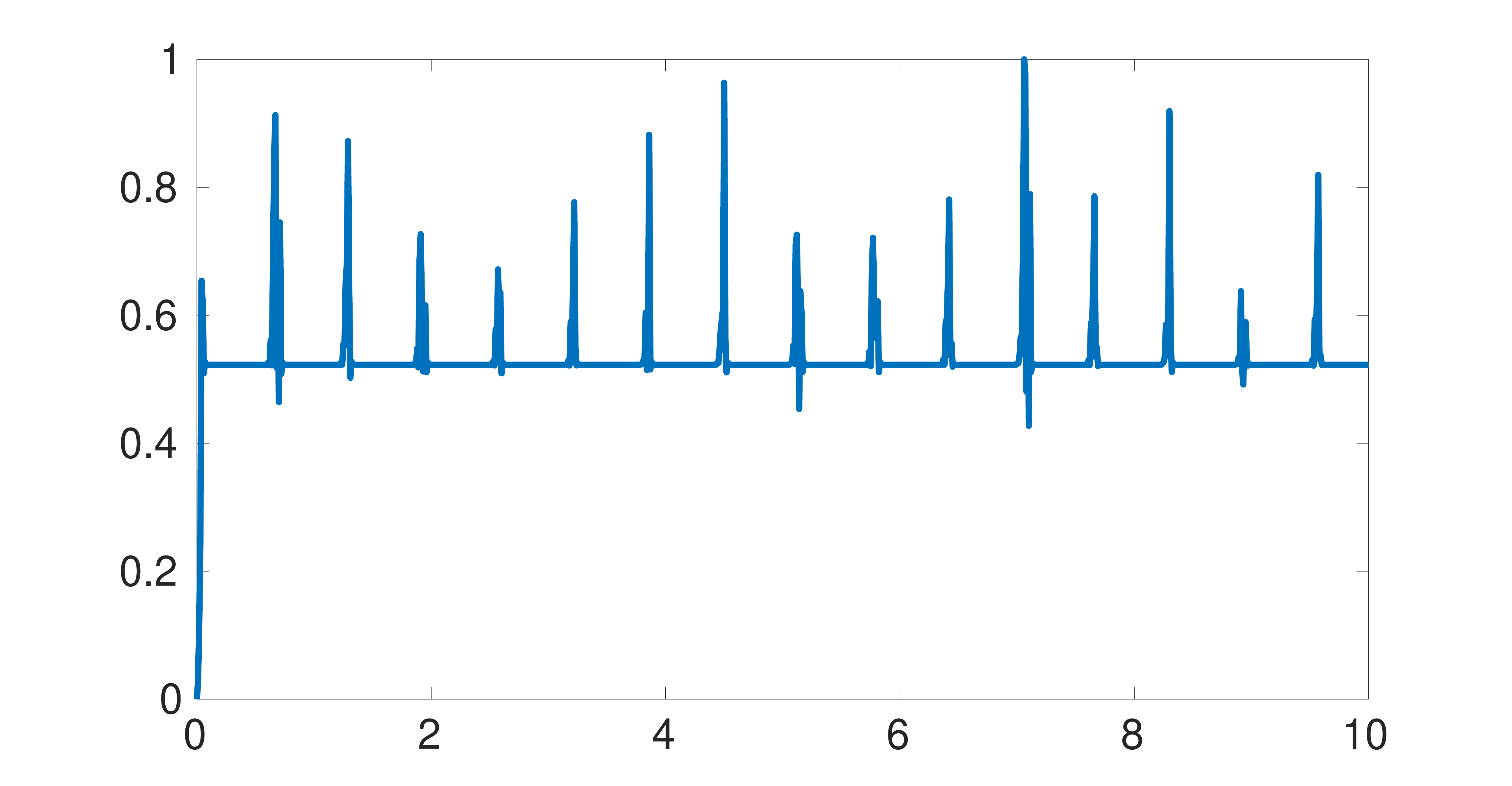}
\put(50,0){$t$}
\put(-1,26){$\mathcal{E}(t)$}
\put(0,50){(f)}
\end{overpic}
\end{minipage}\\
\vspace{0.2cm}
\begin{minipage}{0.49\textwidth} 
\begin{overpic}[width=\textwidth,trim={40 0 0 40},clip]{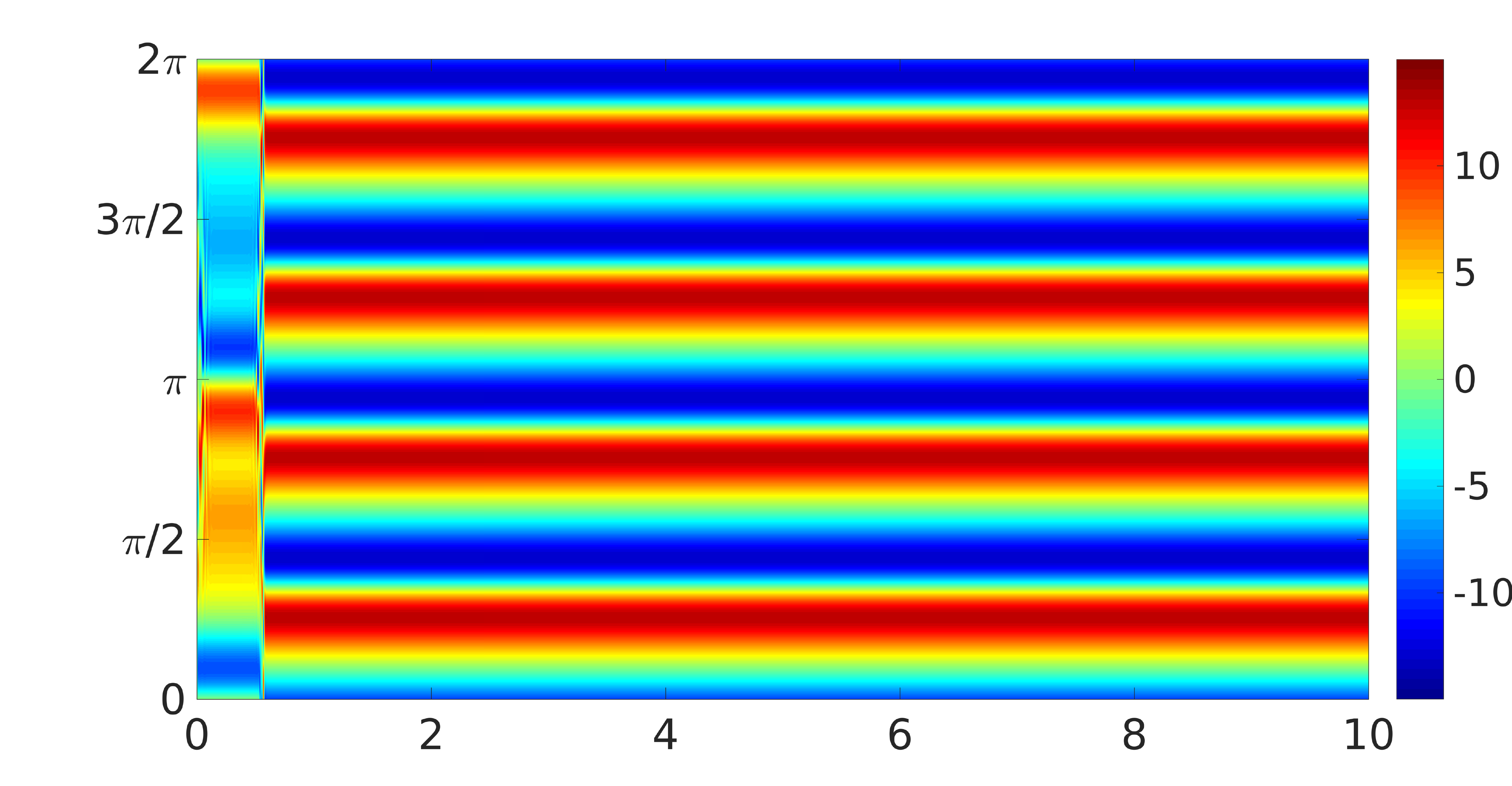}
\put(50,0){$t$}
\put(1,26){$x$}
\put(-2,50){(c)}
\end{overpic}
\end{minipage}
\begin{minipage}{0.49\textwidth} 
\begin{overpic}[width=\textwidth,trim={0 0 40 40},clip]{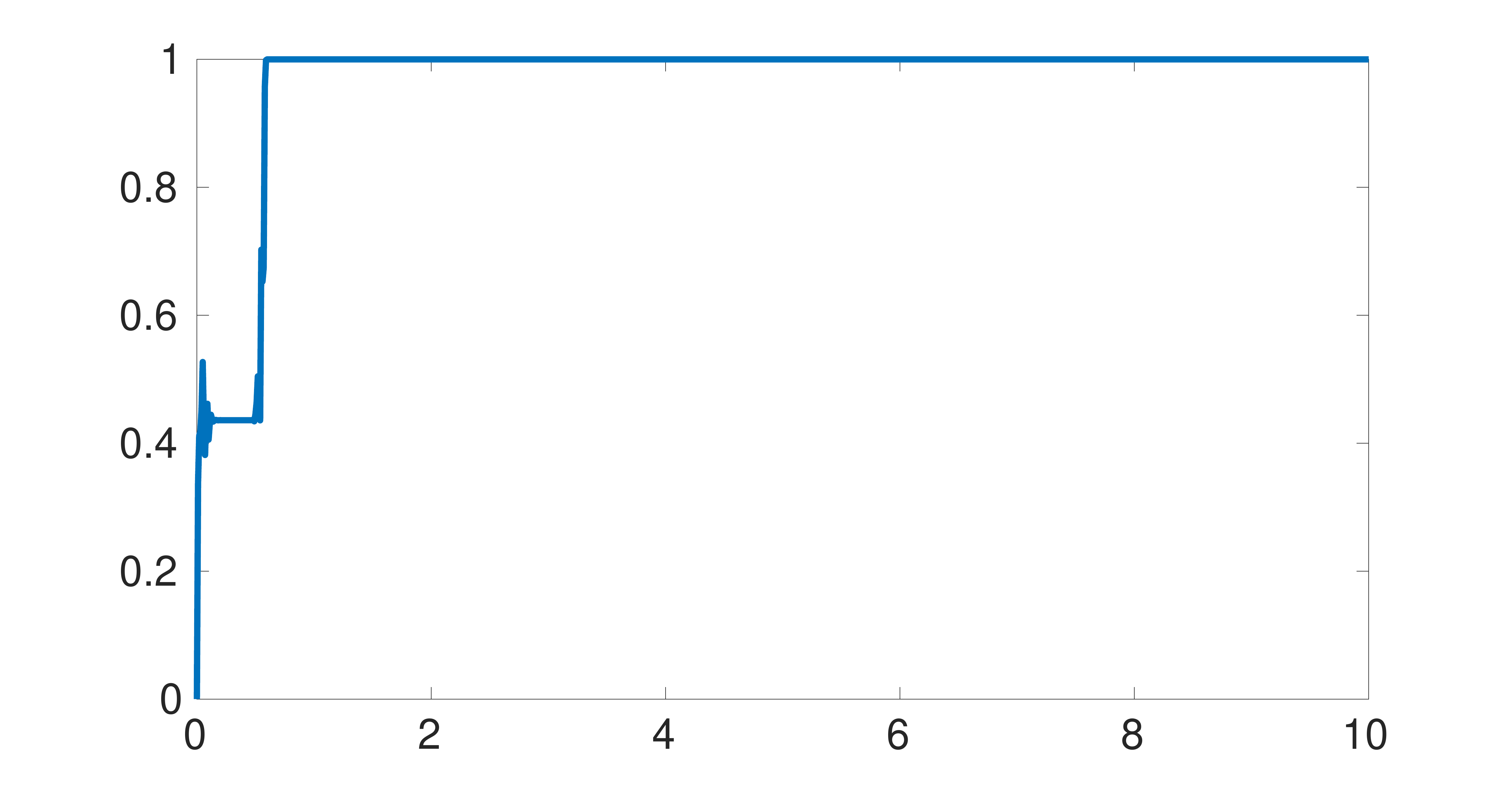}
\put(50,0){$t$}
\put(-1,26){$\mathcal{E}(t)$}
\put(0,50){(g)}
\end{overpic}
\end{minipage}\\
\vspace{0.2cm}
\begin{minipage}{0.49\textwidth} 
\begin{overpic}[width=\textwidth,trim={40 0 0 40},clip]{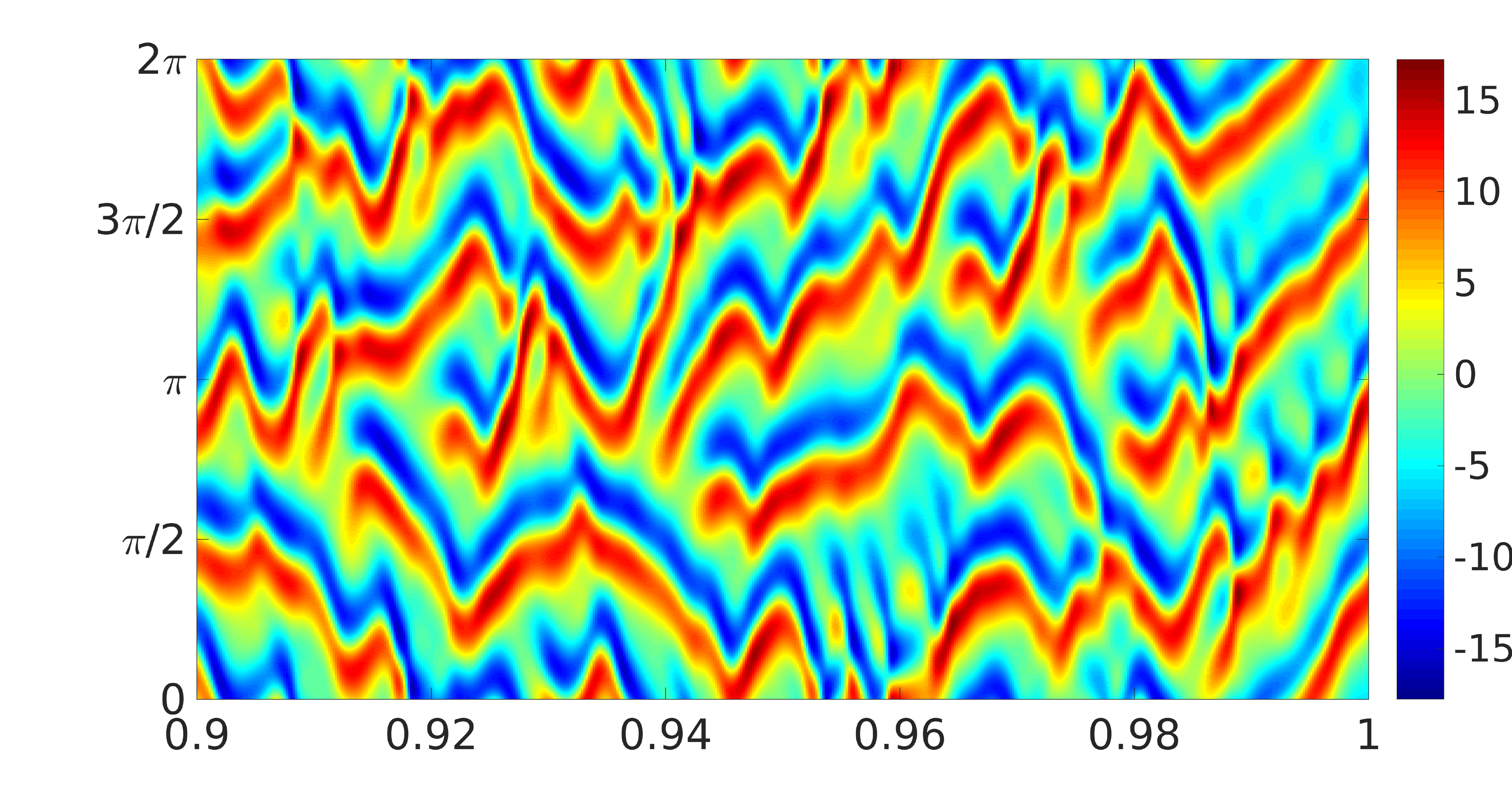}
\put(50,0){$t$}
\put(1,26){$x$}
\put(-2,50){(d)}
\end{overpic}
\end{minipage}
\begin{minipage}{0.49\textwidth} 
\begin{overpic}[width=\textwidth,trim={0 0 40 40},clip]{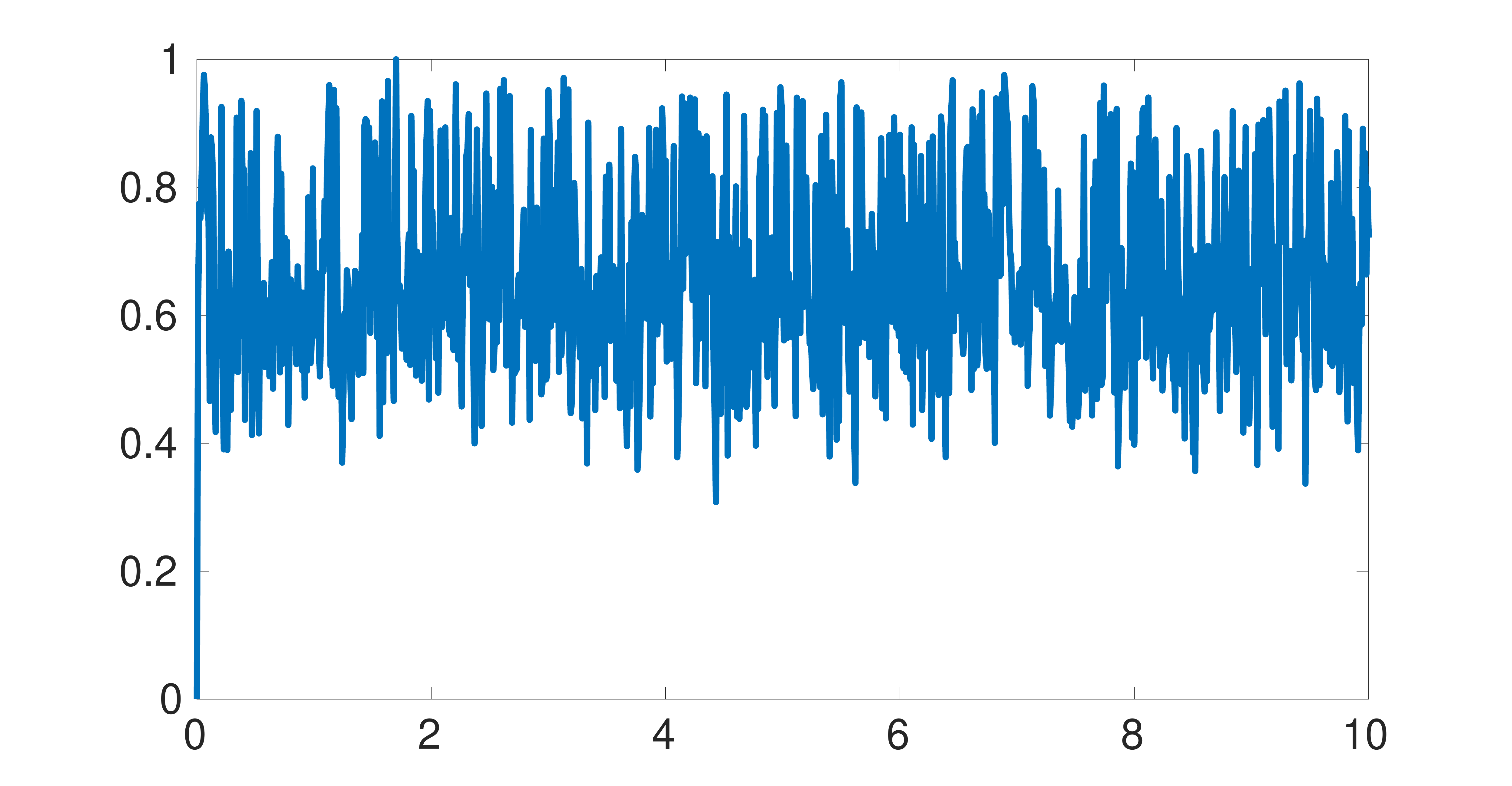}
\put(50,0){$t$}
\put(-1,26){$\mathcal{E}(t)$}
\put(0,50){(h)}
\end{overpic}
\end{minipage}
\caption{Solutions to the KS equation with $\alpha=20$ (a), $44$ (b), $100$ (c), and $125$ (d). The right panels show the corresponding normalized energy $\mathcal{E}(t)$. The chaotic solution depicted in (d) has been zoomed to $t\in[0.9,1]$. (c) and (g) illustrate the transient regime of the solution for $t\in[0,0.5]$ before convergence to the global quadrimodal attractor.}
\label{fig_sol_KS}
\end{figure*}
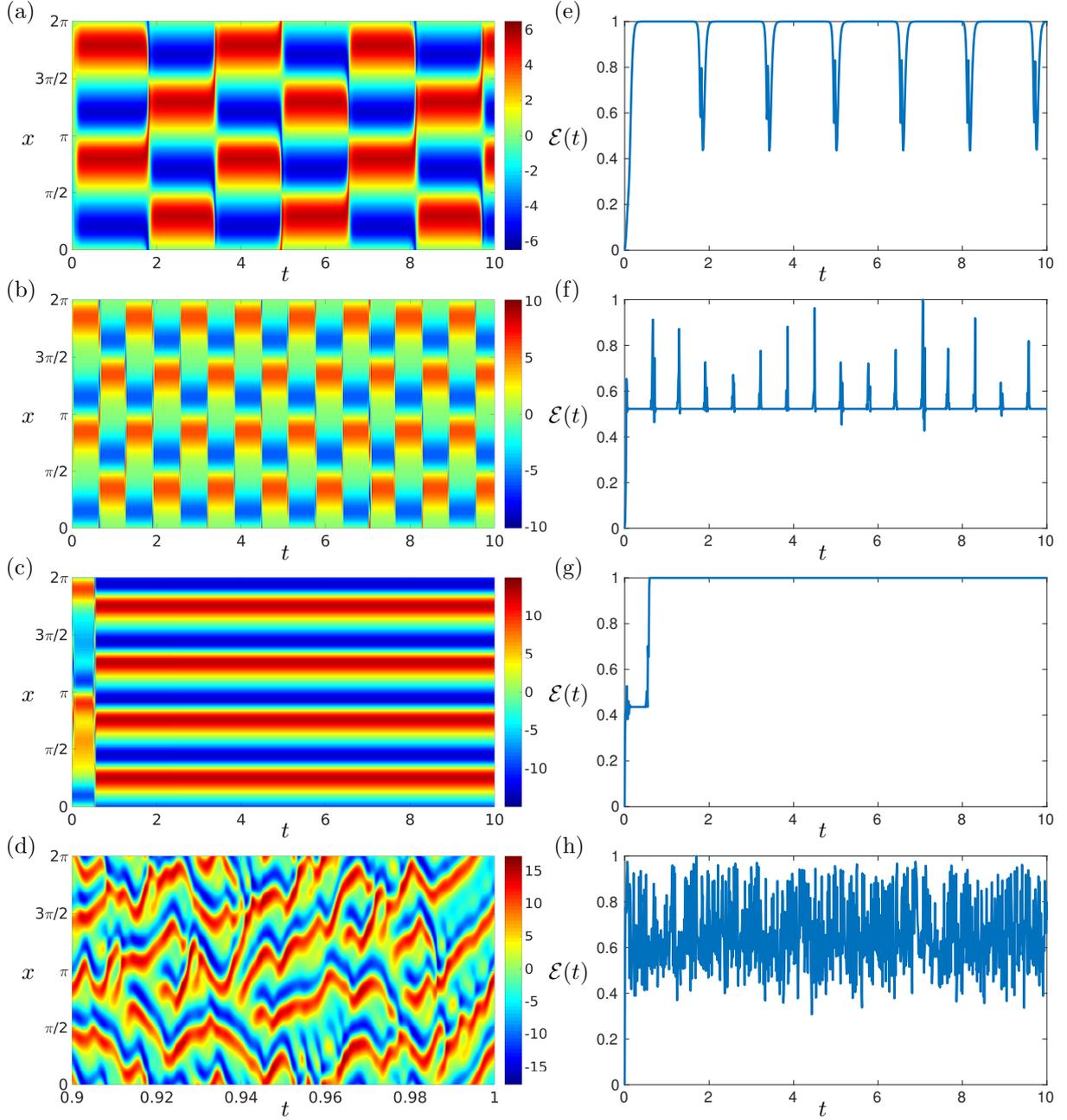

\cref{fig_sol_KS} shows oscillatory solutions to the KS equation for $\alpha = 20$ (a), $44$ (b) and a quadrimodal solution (c). A chaotic solution to the KS equation is depicted in \cref{fig_sol_KS} (d). This spatiotemporal chaotic behaviour is hard to analyse because of the high dimensionality of the system, i.e. the large number of Fourier modes of the solutions. Thus, we analyse the behaviour of the solutions by considering the energy time series
\begin{equation}
\mathcal{E}(t) = \int_{0}^{2\pi}u(x,t)^2\,dx,
\end{equation}
normalized by the transformation $\mathcal{E}(t)\mapsto (\mathcal{E}(t)-m)/(M-m)$, where
\begin{align*}
m &= \min\limits_{t\in [0,10]}\int_{0}^{2\pi}u(x,t)^2\,dx,\\
M &= \max\limits_{t\in [0,10]}\int_{0}^{2\pi}u(x,t)^2\,dx,
\end{align*}
such that it lies in the interval $[0,1]$. The normalized energy time series of the solutions to the KS equation for $\alpha = 20$, $44$, $100$, and $125$ is plotted in \cref{fig_sol_KS} (e) to (h), respectively. These figures illustrate the relation between $\mathcal{E}(t)$ and the behaviour of the solution $u(x,t)$.

Similarly to \cref{sec_lorenz}, we train the large kernel convolutional neural network described in \cref{sec_convnet} on the $x$ component of the Lorenz system and test it on the time series from the data set of the KS equation described above. The global accuracy that we obtain to classify the time series between chaotic and non-chaotic is $94.4\%$. The accuracy for the different classes of attracting solutions in the testing set is reported in \cref{tab_KS_result}.

\begin{table}[htbp]
\centering
\caption{Classification results of the energy time series of the KS equation for various $\alpha$. The neural network LKCNN is trained on the $x$ component of the Lorenz system. The classification accuracy is reported for the different intervals of $\alpha$ composing the data set and averaged over five training cycles.}
\label{tab_KS_result}
\vspace{0.2cm}
\begin{tabular}{lll}
\hline
Range of $\alpha$ & Solutions behaviour & Accuracy \\
\hline
$[18,22]$ & periodic & $54.8$ \\
$[23,33]$ & bimodal & $96.4$ \\
$[43,45]$ & periodic & $99.2$ \\
$[56,65]$ & trimodal & $95.4$ \\
$[95,115]$ & quadrimodal & $99.6$ \\
$[120,130]$ & chaotic & $99.8$ \\
\hline
\end{tabular}
\end{table}

We observe that the LKCNN classifies correctly time series of bimodal, highly oscillatory, trimodal, and quadrimodal solutions, corresponding to $\alpha\in[23,33]$, $[43,45]$, $[56,65]$, $[95,115]$, as non-chaotic with an accuracy of $96.4\%$, $99.2\%$, $95.4\%$, and $99.6\%$, respectively. Moreover, the network achieves $99.8\%$ accuracy on the set of chaotic time series. However, the energy time series of low-frequency periodic solutions to the Kuramoto--Sivashinsky equation for $\alpha\in [18,22]$ are misclassified by the neural network since only $54.8\%$ of them are identified as non-chaotic. We expect this misclassification to be due to qualitative differences between the corresponding energy time series of the KS equation and the periodic time series of the Lorenz system. In particular, the KS data set contains periodic time series with low frequency oscillations in this regime (see \cref{fig_sol_KS} (b)), while the Lorenz system generates periodic time series with high frequency oscillations (see \cref{fig_lorenz_sequence} (c)). The neural network is then unable to classify features that are not present in the training set and hence fails to generalise to the low frequency periodic time series of the Kuramoto--Sivashinsky equation. 

However, we identified two regimes (i.e. $\rho\in[140,160]$ and $\rho\in[200,220]$) for which the $z$ component time series of the Lorenz system are low frequency. Therefore, we trained LKCNN on the $z$ component of the Lorenz system over the whole range of parameters $\rho\in[0,250]$ and tested it on the KS system. Then, we find that the low frequency time series of the KS system (corresponding to $\alpha\in [18,22]$) are now classified correctly with an accuracy of $95.6\%$, (instead of $54.8\%$ when the network is trained on the $x$ component of the Lorenz system). Moreover, the overall accuracy improves to $98.7\%$ instead of $94.4\%$.

\subsection{Accuracy dependence on the training data set size and the time series length}

We test the robustness of the neural network LKCNN on the classification problem of the KS equation by studying how the accuracy depends on the size of the training data set and the length of the time series. Remember that the network is trained on time series of the $x$ component of the Lorenz system of same length, whose chaotic classification is obtained using the Lyapunov exponent.

\begin{figure}[htbp]
\centering
\begin{overpic}[width=0.49\textwidth]{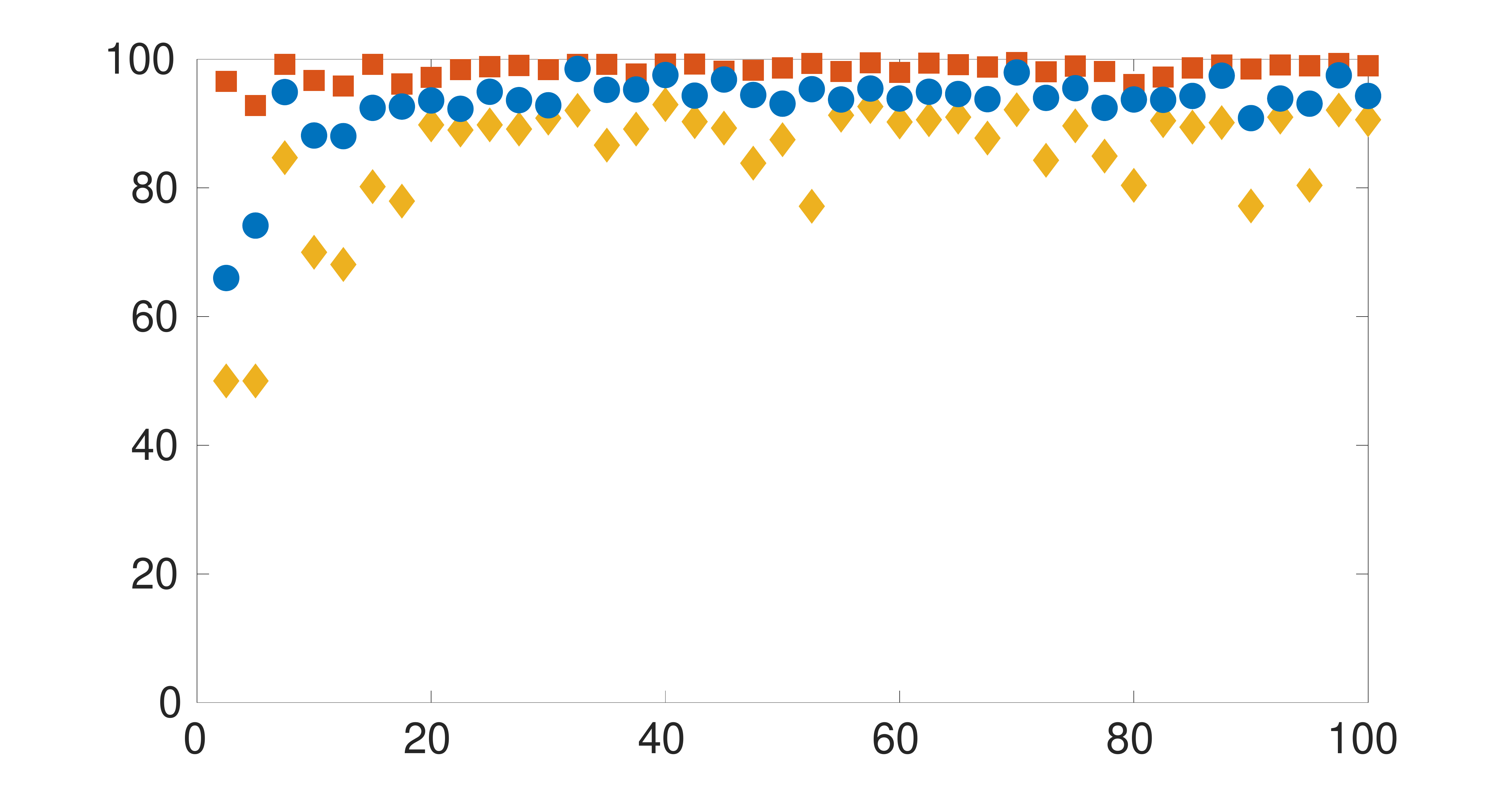}
\put(2,18){\rotatebox{90}{Accuracy}}
\put(24,-3){Amount of training data (\%)}
\end{overpic}
\caption{Classification accuracy of the KS data set versus the amount of training data (percentage from five thousand realisations). We show the first quartile ($\Diamond$), the median ($\Circle$), and the third quartile ($\square$) of the accuracy obtained by $20$ training cycles of the LKCNN neural network on time series of the $x$ component of the Lorenz system.}
\label{fig_robust1}
\end{figure}

\cref{fig_robust1} shows the effect of the size of the training data set on the classification accuracy of LKCNN. 
We observe that the neural network achieves a median accuracy greater than $90\%$ when the amount of training data available is above $10\%$. Note that this $10\%$ corresponds to five hundred time series from the whole data set of the Lorenz system. This high accuracy is due to the accurate classification of the neural network on most regimes of the time series of the data set (see \cref{tab_KS_result}). It is of interest to notice that the accuracy is slightly affected only for the first quartile above $10\%$ of the size of the training set. The fluctuations of the accuracy in \cref{fig_robust1} are explained by the difficulty to classify low frequency time series of the KS data set.

\begin{figure}[htbp]
\centering
\begin{overpic}[width=0.49\textwidth]{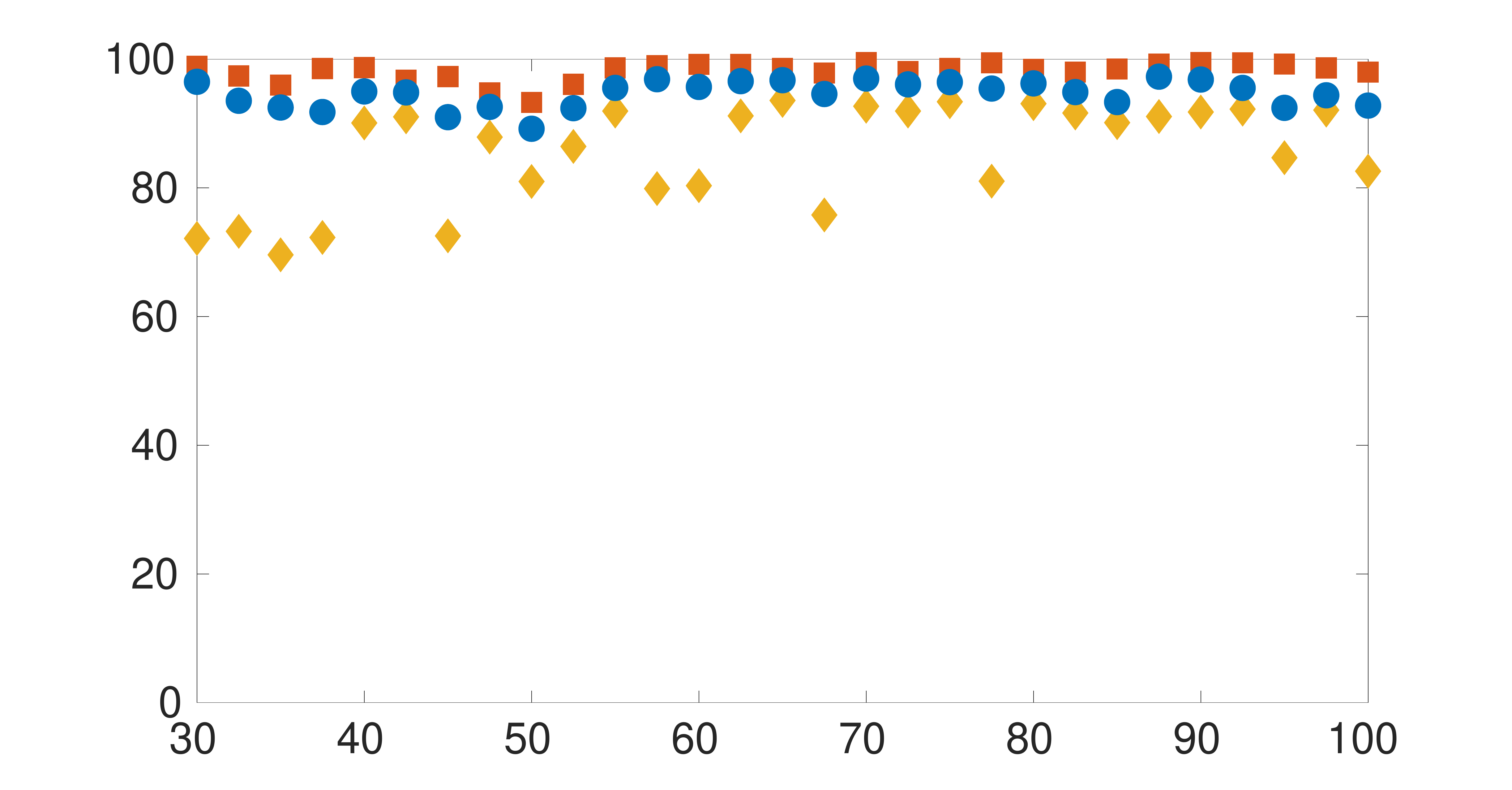}
\put(2,18){\rotatebox{90}{Accuracy}}
\put(24,-3){Length of the time series (\%)}
\end{overpic}
\caption{Classification accuracy of the KS time series with respect to the length of the time series (percentage from one thousand timesteps). We show the first quartile ($\Diamond$), the median ($\Circle$), and the third quartile ($\square$) of the accuracy obtained by $20$ training cycles of the LKCNN neural network on time series of the $x$ component of the Lorenz system.}
\label{fig_robust2}
\end{figure}

In \cref{fig_robust2}, we analyse the classification ability of LKCNN on shorter time series. It is interesting that for all the lengths of the time series we considered, LKCNN reaches a median accuracy greater than $90\%$ on the KS problem. This indicates that the sign of the Lyapunov exponent even for short time series of length three hundred of the training set is determined as accurate as for time series of length thousand. Consequently, this result suggests that the Lyapunov exponent is a good measure of chaotic behaviour in the Lorenz system. The fact that we train and test the network on time series of the same length might also be a justification of this high accuracy. On the other hand, the fluctuations of the first quartile show the inability of the network to classify the time series of the KS data set in a few cases. Note that we have not considered time series of lengths less than three hundred (or 30\% of the time series length). This is because LKCNN includes two convolutional layers of kernel size one hundred.

Overall, our results show the robustness of LKCNN on this problem. In particular, this network is able to generalise well on time series generated by the KS equation and achieves a median classification accuracy greater than $90\%$, independently of the size of the training data set or the length of the time series.

\section*{Conclusions}
\addcontentsline{toc}{section}{Conclusions}

For some of the most challenging real-life applications the expression of a precise underlying dynamical system is unknown or the phase space of the system is infinite dimensional, which makes the identification of the different dynamical regimes unfeasible or in the best case scenario computationally expensive. For this reason, in this study we have introduced a deep learning approach for classifying univariate time series generated by discrete and continuous dynamical systems. Our approach is to train standard neural networks on a given dynamical system with a basic or low-dimensional phase space and generalise by using this network to classify univariate time series of a dynamical system with more intricate or high-dimensional phase space.

The convolutional neural network with large kernel size (LKCNN) is able to learn the chaotic features of these dynamical systems and classify their time series with high accuracy. On the other hand, state-of-the-art neural networks tend to overfit the training data set and do not perform as well. In detail, our generalisation approach has been applied to classify univariate time series generated by the sine-circle map and the Kuramoto--Sivashinsky equation, using the logistic map and the Lorenz system as training data sets, respectively. We observed a classification accuracy greater than $80\%$ on the sine-circle map and more than $90\%$ on the KS equation. This great performance can be justified from the fact that there are no batch normalisation layers in this network and thus we avoid rescaling the testing data with the wrong normalisation parameters in contrast with FCN and ResNet. Moreover, we observed that the classification accuracy of the KS time series increased further when low frequency time series were involved in the training phase. We also found that LKCNN is able to generalise with high classification accuracy even when the size of the training data set or the length of the time series is very small. This demonstrates the robustness of this neural network in this problem.

Finally, our study suggests that deep learning techniques, which can generalise the knowledge acquired from a training set to a different testing set, can be valuable to classify time series into chaotic and non-chaotic obtained from real-life applications. There are many directions in which the present results can be pursued further. First of all, attempting to classify time series obtained from real-life applications is crucial. In that respect, the effect of noise in the training and testing data sets is an important aspect to be considered and study the influence of the noise to the accuracy of the networks to classify the time series. We hope to address such an analysis in our future work.

\section*{Acknowledgments}

The authors would like to thank Samuel Cohen for useful discussions and Nicos Pavlidis for providing useful comments on an early draft of this manuscript. We thank the two anonymous reviewers for their valuable comments and suggestions that improved our manuscript. This work was supported by the EPSRC Centre For Doctoral Training in Industrially Focused Mathematical Modelling (EP/L015803/1) in collaboration with Culham Centre for Fusion Energy. This work has been carried out within the framework of the EUROfusion Consortium and has received funding from the Euratom research and training programme 2014-2018 and 2019-2020 under grant agreement No 633053. The views and opinions expressed herein do not necessarily reflect those of the European Commission.

\appendix
\section{Sensitivity of LKCNN's performance to the kernel size and feature channels}
\label{appendix_a}

We vary the size of the convolutional kernel and number of feature channels in the large kernel convolutional neural network (LKCNN) to study the sensitivity of the classification accuracy with respect to these parameters. LKCNN is trained on the logistic signals and tested on signals generated by the sine-circle map (see \cref{sec_discrete_res}).

\begin{figure}[htbp]
\centering
\begin{overpic}[width=0.49\textwidth]{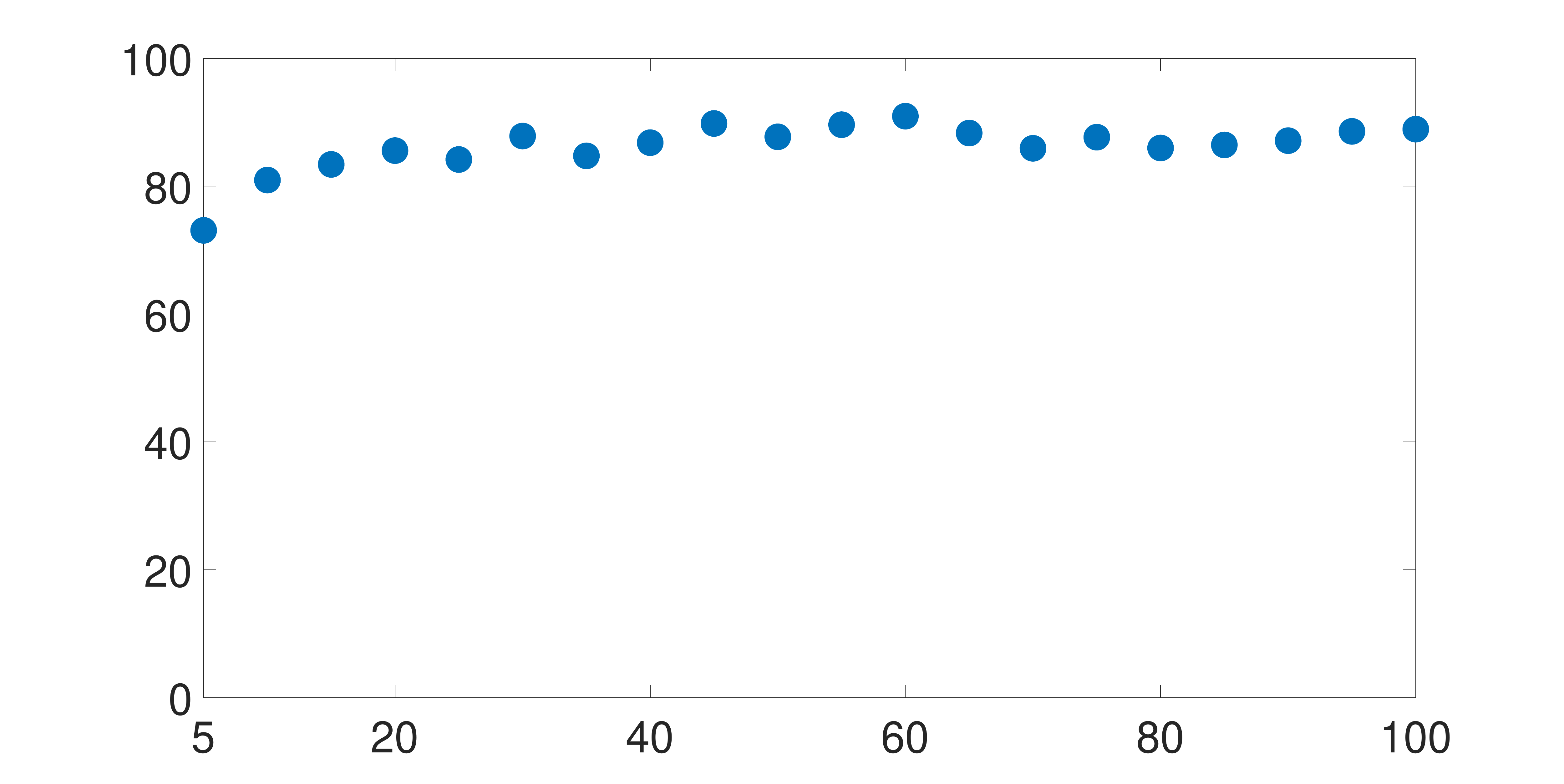}
\put(2,18){\rotatebox{90}{Accuracy}}
\put(42,-3){Kernel size}
\end{overpic}
\caption{Classification accuracy of the sine-circle map signals with respect to the convolutional kernel size.  LKCNN is trained on logistic signals with five feature channels and the accuracy is averaged over five training cycles.}
\label{fig_A_1}
\end{figure}

In \cref{fig_A_1}, we fix the number of feature channels to five and vary the size of the convolutional kernel from five to a hundred. We obtain a classification accuracy between $80\%$ and $90\%$. This indicates that the network is not very sensitive to the size of the convolutional kernel, which is usually chosen to be small.

\begin{figure}[htbp]
\centering
\begin{overpic}[width=0.49\textwidth]{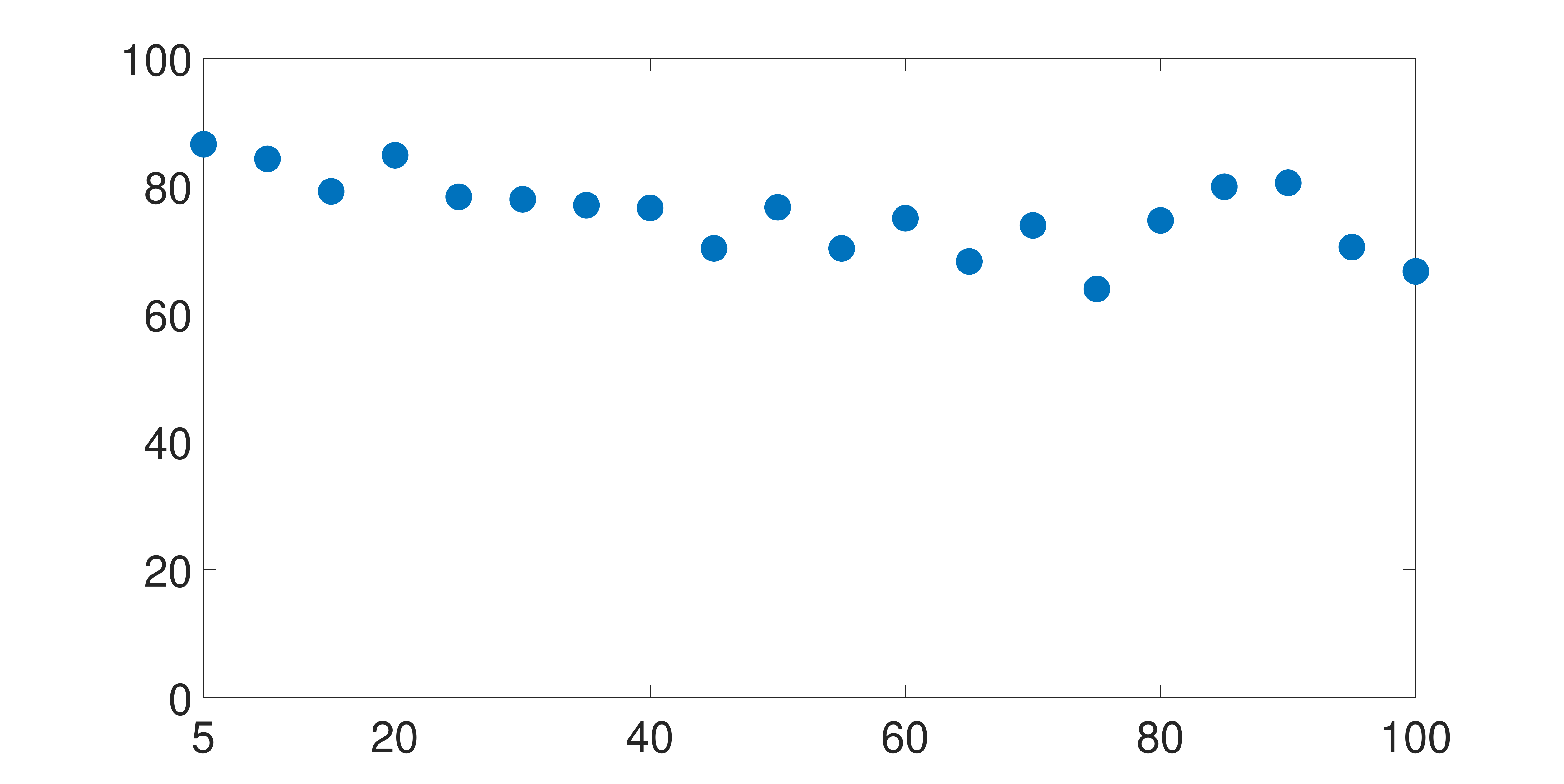}
\put(2,18){\rotatebox{90}{Accuracy}}
\put(26,-3){Number of feature channels}
\end{overpic}
\caption{Classification accuracy of the sine-circle map signals with respect to the number of feature channels. LKCNN is trained on logistic signals with a kernel size of a hundred and the accuracy is averaged over five training cycles.}
\label{fig_A_2}
\end{figure}

Now, we analyse the dependence of the classification accuracy of LKCNN in \cref{fig_A_2} by keeping the kernel size fixed to a hundred and varying the number of feature channels. We observe that the accuracy decreases from $85\%$ to approximatively $65\%$ when the number of channels increases. An explanation of this behaviour is that the number of parameters for LKCNN dramatically increases as we increase the number of feature channels. This makes the training phase computationally more challenging because the optimisation algorithm struggles to converge due to the small number of epochs that we have chosen (see \cref{sec_Net_TSC}).

Based on these results, we chose to have large convolutional kernels to decrease the computational expense of the network and small number of feature channels to avoid overfitting the data. With these choices we improve the generalisation ability and the computational performance of the network.

%\section*{\refname} % Comment this if you have double reference
\bibliography{biblio}

\end{document}